\RequirePackage{fix-cm}
\documentclass[twocolumn,epjc3]{svjour3}  
\smartqed  
\RequirePackage{graphicx}
\usepackage[version=4]{mhchem}
\usepackage[T1]{fontenc}    
\usepackage[utf8]{inputenc}    
\usepackage[english]{babel}
\usepackage{comment}
\usepackage{lipsum}
\usepackage{bm}
\usepackage{graphicx}
\usepackage{subfig}
\usepackage{siunitx}
\usepackage{caption}
\usepackage{microtype}

\journalname{Eur. Phys. J. C}

\begin{document}

\title {\ce{^{222}Rn} contamination mechanisms on acrylic surfaces}

\author{M. Nastasi\thanksref{addr1,addr2}
       \and A.~Paonessa\thanksref{addr1} 
       \and E.~Previtali\thanksref{addr1,addr2}
       \and E.~Quadrivi\thanksref{addr1,addr19,e1} 
       \and M.~Sisti\thanksref{addr1,addr2} 
       \and S.~Aiello\thanksref{addr3,addr4}
       \and G.~Andronico\thanksref{addr3,addr4}
       \and V.~Antonelli\thanksref{addr5,addr6}
       \and W.~Baldini\thanksref{addr17,addr18}
       \and M.~Bellato\thanksref{addr8}
       \and A.~Bergnoli\thanksref{addr7,addr8}
       \and A.~Brigatti\thanksref{addr5,addr6}
       \and R.~Brugnera\thanksref{addr7,addr8}
       \and A.~Budano\thanksref{addr9,addr10}
       \and M.~Buscemi\thanksref{addr3,addr4}
       \and A.~Cammi\thanksref{addr11,addr2} 
       \and R.~Caruso\thanksref{addr3,addr4}
       \and D.~Chiesa\thanksref{addr1,addr2}
       \and C.~Clementi\thanksref{addr12,addr13}
       \and D.~Corti\thanksref{addr8}
       \and S.~Costa\thanksref{addr3,addr4}
       \and F.~Dal Corso\thanksref{addr8}
       \and X.F.~Ding\thanksref{addr15,addr5,addr6}
       \and S.~Dusini\thanksref{addr8}
       \and A.~Fabbri\thanksref{addr9,addr10}
       \and G.~Fiorentini\thanksref{addr17,addr18}
       \and R.~Ford\thanksref{addr16,addr5,addr6}
       \and A.~Formozov\thanksref{addr5,addr6}
       \and G.~Galet\thanksref{addr7,addr8}
       \and A.~Garfagnini\thanksref{addr7,addr8}
       \and M.~Giammarchi\thanksref{addr5,addr6}
       \and A.~Giaz\thanksref{addr7,addr8}
       \and M.~Grassi\thanksref{addr5,addr6}
       \and R.~Isocrate\thanksref{addr8}
       \and C.~Landini\thanksref{addr5,addr6}
       \and I.~Lippi\thanksref{addr8}
       \and P.~Lombardi\thanksref{addr5,addr6}
       \and Y.~Malyshkin\thanksref{addr10}
       \and F.~Mantovani\thanksref{addr17,addr18}
       \and S.M.~Mari\thanksref{addr9,addr10}
       \and F.~Marini\thanksref{addr7,addr8}
       \and C.~Martellini\thanksref{addr9,addr10}
       \and A.~Martini\thanksref{addr14}
       \and E.~Meroni\thanksref{addr5,addr6}
       \and M.~Mezzetto\thanksref{addr8}
       \and L.~Miramonti\thanksref{addr5,addr6}
       \and P.~Montini\thanksref{addr9,addr10}
       \and M.~Montuschi\thanksref{addr17,addr18}
       \and F.~Ortica\thanksref{addr12,addr13}
       \and A.~Paoloni\thanksref{addr14}
       \and S.~Parmeggiano\thanksref{addr5,addr6}
       \and N.~Pelliccia\thanksref{addr12,addr13}
       \and G.~Ranucci\thanksref{addr5,addr6}
       \and A.C.~Re\thanksref{addr5,addr6}
       \and B.~Ricci\thanksref{addr17,addr18}
       \and A.~Romani\thanksref{addr12,addr13}
       \and P.~Saggese\thanksref{addr5,addr6}
       \and G.~Salamanna\thanksref{addr9,addr10}
       \and F.H.~Sawi\thanksref{addr7,addr8}
       \and A.~Serafini\thanksref{addr17,addr18}
       \and G.~Settanta\thanksref{addr9,addr10}
       \and C.~Sirignano\thanksref{addr7,addr8}
       \and L.~Stanco\thanksref{addr8}
       \and V.~Strati\thanksref{addr17,addr18}
       \and C.~Tuv\`e\thanksref{addr3,addr4}
       \and G.~Verde\thanksref{addr3,addr4}
       \and L.~Votano\thanksref{addr14}
 }
 
\thankstext{e1}{e-mail: eleonora.quadrivi@eni.com}


\institute{Dipartimento di Fisica, Universit\`{a} degli Studi di Milano-Bicocca, 
          I-20126 Milano, Italy  \label{addr1}
          \and
          Istituto Nazionale di Fisica Nucleare, Sezione Milano-Bicocca, 
          I-20126 Milano, Italy \label{addr2}
          \and
          Istituto Nazionale di Fisica Nucleare, Sezione di Catania, 
          I-95123 Catania, Italy \label{addr3}
          \and
          Dipartimento di Fisica e Astronomia, Universit\`{a} di Catania, 
          I-95123 Catania, Italy \label{addr4}
          \and Dipartimento di Fisica, Universit\`{a} di Milano, 
          I-20133 Milano, Italy \label{addr5}
          \and
          Istituto Nazionale di Fisica Nucleare, Sezione di Milano, 
          I-20133 Milano, Italy   \label{addr6}    
          \and
          Istituto Nazionale di Fisica Nucleare, Sezione di Ferrara,
          I-44122 Ferrara, Italy \label{addr17}
          \and
          Dipartimento di Fisica e Scienze della Terra, Universit\`{a} di Ferrara, 
          I-44122 Ferrara, Italy \label{addr18}
          \and
          Dipartimento di Fisica e Astronomia, Universit\`{a} di Padova, 
          I-35131 Padova, Italy \label{addr7}
          \and
          Istituto Nazionale di Fisica Nucleare, Sezione di Padova, 
          I-35131 Padova, Italy \label{addr8}
          \and
          Dipartimento di Matematica e Fisica, Universit\`{a} di Roma Tre, 
          I-00146 Roma, Italy \label{addr9}
          \and
          Istituto Nazionale di Fisica Nucleare, Sezione di Roma Tre, 
          I-00146 Roma, Italy \label{addr10}
          \and
           Politecnico di Milano, Dipartimento di Energia, 
           I-20156, Milano, Italy \label{addr11}
           \and
           Dipartimento di Chimica, Biologia e Biotecnologie, Universit\`{a} di Perugia, 
           I-06123 Perugia, Italy \label{addr12}
           \and
           Istituto Nazionale di Fisica Nucleare, Sezione di Perugia,
           I-06123 Perugia, Italy \label{addr13}
           \and
           Istituto Nazionale di Fisica Nucleare, Laboratori Nazionali di Frascati, 
           I-00044 Frascati (RM), Italy \label{addr14}
           \and
           Gran Sasso Science Institute, 
           I-67100 L'Aquila (AQ), Italy   \label{addr15}
           \and
           SNOLAB, 
           Lively, ON P3Y 1N2, Canada \label{addr16}
           \and
           \emph{Present Address: Centro Ricerche Eni Bolgiano, via Felice Maritano 24, I-20097 San Donato Milanese (MI), Italy} \label{addr19}
}

\date{Received: date / Accepted: date}

\maketitle

\begin{abstract}

In this work, the \ce{^{222}Rn} contamination mechanisms on acrylic surfaces have been investigated. \ce{^{222}Rn} can represent a significant background source for low-background experiments, and acrylic is a suitable material for detector design thanks to its purity and transparency. Four acrylic samples have been exposed to a \ce{^{222}Rn} rich environment for different time periods, being contaminated by \ce{^{222}Rn} and its progenies. Subsequently, the time evolution of radiocontaminants activity on the samples has been evaluated with $\alpha$ and $\gamma$ measurements, highlighting the role of different decay modes in the contamination process. A detailed analysis of the alpha spectra allowed to quantify the implantation depth of the contaminants. Moreover, a study of both $\alpha$ and $\gamma$ measurements pointed out the \ce{^{222}Rn} diffusion inside the samples.

\end{abstract}

\section{Introduction}
\label{intro}

\begin{figure}
\includegraphics[scale = 0.37]{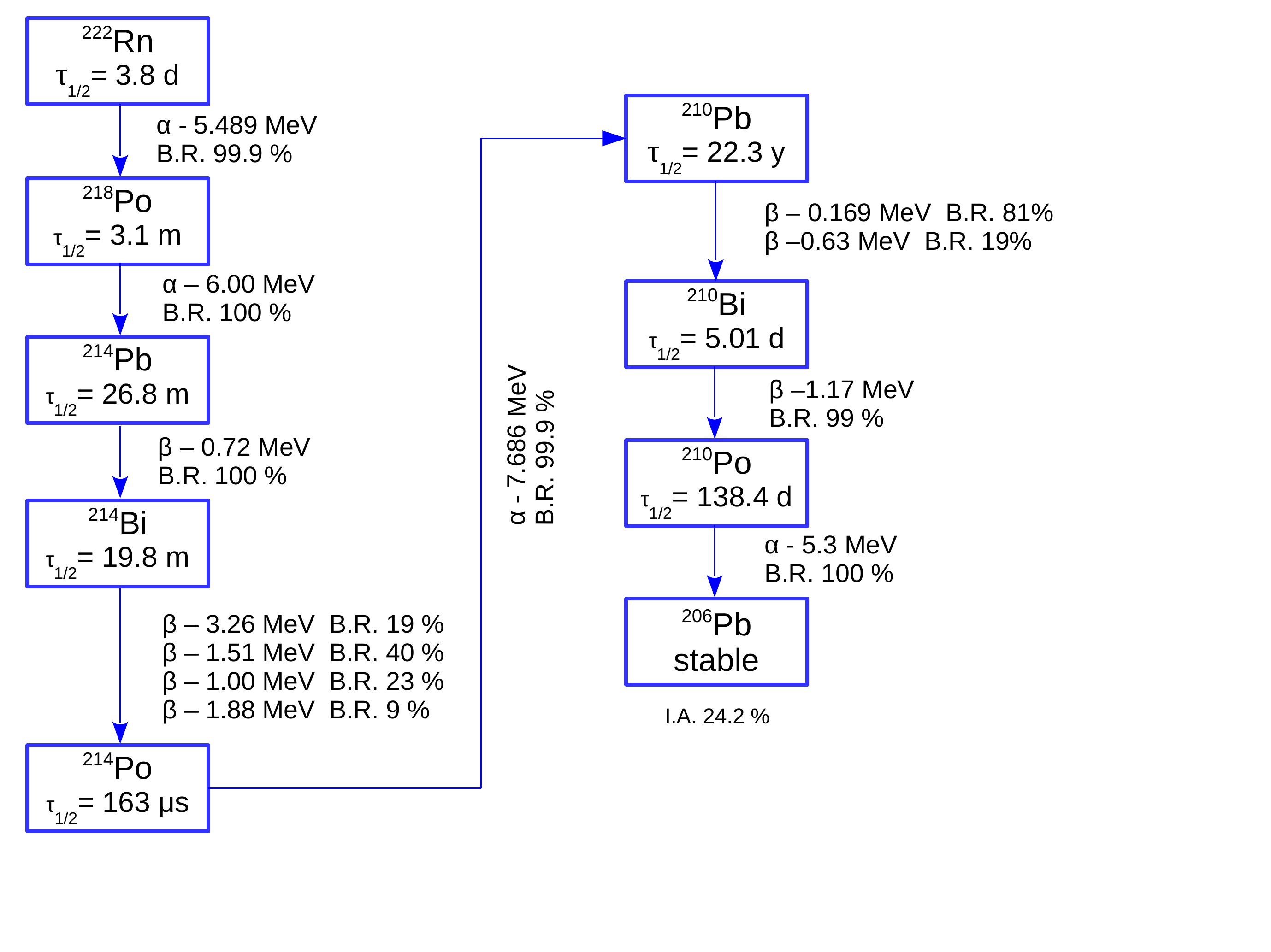}
\caption{The \ce{^{222}Rn} chain. All the $\gamma$ decays have been omitted.}
\label{fig:1}
\end{figure}

Radon-$222$ is a radioactive noble gas belonging to the \ce{^{238}U} chain, a natural chain that is present in almost all the rock types. Thanks to its chemical inertia and its half-life of $3.8$ days, this gas can diffuse through rock layers containing uranium ore and spread in air. Another way for \ce{^{222}Rn} to reach the atmosphere is the dissolution of its parent \ce{^{226}Ra} in water which filters through rocks. Both these mechanisms make possible to consider a \ce{^{222}Rn} chain that is independent from the \ce{^{238}U} chain where \ce{^{222}Rn} originates (Fig.~\ref{fig:1}). When a material is exposed to air, it is naturally exposed to \ce{^{222}Rn}, leading to a radioactive contamination on it.

\ce{^{222}Rn} can contaminate the detection system of an experiment and, for some categories such as rare events searches, it can represent a not negligible source of background. For this reason, understanding the \ce{^{222}Rn} contamination mechanisms is a primary step in the experiment planning, and is also necessary in a wide range of applications, wherever an ultra-pure material is needed. A \ce{^{222}Rn} contamination may occur during the whole material life. Indeed, raw materials can be naturally contaminated by \ce{^{222}Rn} present in mining deposits; a radiopure material can also be exposed and re-contaminated during all construction phases as production, handling and storage. Additionally, in those low-background experiments where the detector is placed underground, a further \ce{^{222}Rn} exposure is likely to occur because of the presence of rocks in the surroundings. In these experiments (e.g., CUORE \cite{Pattavina}, JUNO \cite{JUNO}, SNO+ \cite{SNO}), a study of the \ce{^{222}Rn} related background is a key point to achieve the demanded performances. 

A contamination can take place at each point of the \ce{^{222}Rn} chain. The action of the following three mechanisms is expected: 
\begin{itemize}
\item Diffusion of \ce{^{222}Rn} itself inside materials due to a concentration gradient.
\item Implantation of radioactive \ce{^{222}Rn} progenies as a result of nuclear decays.
\item Deposition on surfaces of dust bound to radioactive particles in air. 
\end{itemize}
\noindent Among \ce{^{222}Rn} daughters (Fig.~\ref{fig:1}), \ce{^{210}Pb} ($\tau_{1/2} = \SI{22.3}{y}$) is the isotope which may be responsible for the longest lasting contamination. This kind of contamination may be not easily removable and a continuous source of background, depending on the application. 

The evolution of \ce{^{222}Rn} chain in time and, more generally, of a nuclear chain with $N_0$ nuclei of type 1 and none of the other types initially present can be described by the \emph{Bateman Equations} \cite{Krane}. In these equations, the activity at time $t$ of the $n$th member of the chain is expressed in terms of the decay constants $\lambda_{i}$ of its predecessors:
\begin{equation}
\label{Bateman}
\begin{split}
A_n & = N_0 \sum_{i=1}^{n} c_i e^{-\lambda_i t}  \\
& = N_0 (c_1 e^{-\lambda_1 t} + c_2 e^{-\lambda_2 t} + \dots + c_n e^{-\lambda_n t})
\end{split} 
\end{equation}
where
\begin{equation}
\label{BatemanCoeff}
\begin{split}
c_m & = \frac{\prod_{i=1}^{n}\lambda_i}{{\prod_{i=1}^{n}}' (\lambda_i - \lambda_m)} \\
& = \frac{\lambda_1 \lambda_2 \lambda_3 \dots \lambda_n}{(\lambda_1 - \lambda_m) (\lambda_2 - \lambda_m) \dots (\lambda_n - \lambda_m)}
\end{split}
\end{equation}
The prime on the lower product indicates that the term $i=m$ has to be omitted.

In this work, acrylic has been chosen as a target to study \ce{^{222}Rn} contamination mechanisms. Acrylic is a good choice when a transparent and ultra-pure plastic is needed. Indeed, it covers a primary role in JUNO and SNO+ experiments, where it has been chosen as the construction material for the vessel containing the liquid scintillator. The vessel is in direct contact with the scintillator, thus any contamination would bring to a background that may dangerously increase the spurious count rate.

\section{Contamination Strategy and Measurements}
\label{sec:1}

\begin{figure}
\includegraphics[scale = 0.45]{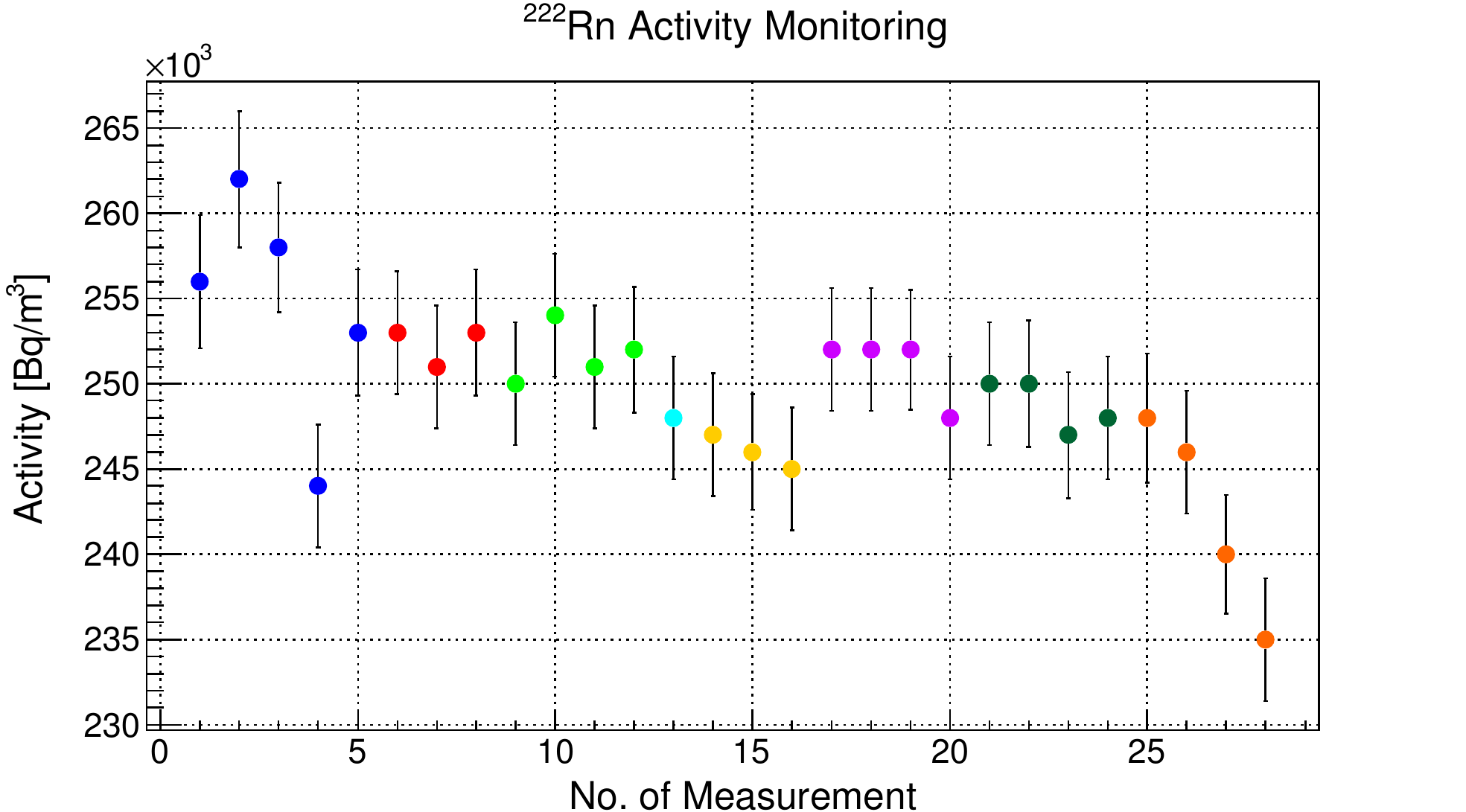}
\caption{\ce{^{222}Rn} activity measurements inside the Rn-Box over a week. Each colour indicates a different day, and points of the same colour correspond to different measurements carried out in a single day.}
\label{fig:RnMon}
\end{figure}

Radon-222 contamination on acrylic has been reproduced in a high radon concentration environment, where the activity of this gas reached values ten thousand times greater than atmospheric levels. This condition enabled to reproduce the consequences of years long exposure of acrylic to air, thus to perform a consistent study in a reasonable time frame. A plexiglass hermetic box (\emph{Rn-Box}) has been employed together with some rocks containing uranium ore, which have been put inside the box following the configuration described in Ref.~\cite{Pattavina}. \ce{^{222}Rn} level inside the Rn-Box has been monitored by daily measurements: in about $20$ days, the Rn-Box reached a \ce{^{222}Rn} saturation level of~$250\pm5$~\si{kBq\per m^3}. In Figure~\ref{fig:RnMon}, monitoring measurements in a week are shown; each point corresponds to a $40$ minutes long measurement, the shortest time frame necessary for the \ce{^{222}Rn} detector to perform a reliable measurement. After the closing of the Rn-Box, all the elements of the \ce{^{222}Rn} chain tend to reach the \emph{secular equilibrium} condition, where all their activities are equal. In a parent-daughter system, secular equilibrium is reached in about five times the daughter half-life. More generally, in a chain case secular equilibrium is reached in about five times the longest living daughter half-life. Hence, in about three hours from the closing ($\approx 5\tau_{1/2}$ of \ce{^{214}Pb}, the longest living daughter before \ce{^{210}Pb}), the upper part of the \ce{^{222}Rn} chain reached the secular equilibrium condition, that is $A_{^{222}Rn} = A_{^{218}Po} = A_{^{214}Pb} = A_{^{214}Bi} = A_{^{214}Po}$. Because of its half-life of~\SI{22.3}{y}, \ce{^{210}Pb} can be considered as an almost stable element that accumulates during the exposure periods. The activity of its daughter (\ce{^{210}Po}) increases towards the condition $A_{^{210}Po} = A_{^{210}Pb}$, which, in our case, is never achieved during the exposure.
 
Four acrylic plates with dimension of~$5\times5\times0.5$~\si{cm} have been chosen as samples, one with \emph{smooth} texture (i.e., a polished surface) and three with \emph{opaque} texture (i.e., a rough surface). After a cleaning process with distilled \ce{H_{2}O}, all the samples have been placed inside the Rn-Box and left in for different periods, as reported in Table~\ref{tab:1}. 

\begin{table}
\caption{Exposure periods of the acrylic samples in order of extraction.}
\label{tab:1} 
\centering
\begin{tabular}{cc}
\hline\noalign{\smallskip}
Sample & Exposure Time \\
\noalign{\smallskip}
	& [d]	\\
\noalign{\smallskip}\hline\noalign{\smallskip}
Smooth  & 86  \\
Opaque 1  & 72 \\
Opaque 2  & 90 \\
Opaque 3  & 39 \\
\noalign{\smallskip}\hline
\end{tabular}
\end{table}

After the exposure, $\alpha$ and $\gamma$ spectroscopies have been performed on each sample. The available detectors were a surface barrier silicon detector with a \SI{900}{mm^2} active surface and a \SI{40}{nm} dead layer placed in a vacuum chamber, for the $\alpha$ spectroscopy, and a broad-energy high-purity germanium detector (HPGe) with carbon window, configured for low-background measurement and with $50\%$ relative efficiency, for the $\gamma$ spectroscopy. At first, a study of the detector background has been carried out. Two two-week-long $\alpha$ measurements have been performed using the silicon detector, one with the chamber being empty and the other with a cleaned but not contaminated acrylic plate. Both the spectra present a compatible activity of \ce{^{210}Po}, at a level of~$(1.0\pm0.1)\times 10^{-4}$~\si{Bq}, negligible with respect to the \ce{^{210}Po} activities obtained for the contaminated samples ($\approx10^{-1}$\,\si{Bq}). Similarly, a study of the HPGe gamma background has shown  a negligible count rate with respect to the measurements of the contaminated samples.

Once a sample was extracted from the Rn-Box for the experimental measurements, a study of the evolution in time of the radon chain element activity was performed in order to understand the \ce{^{222}Rn} contamination features. \ce{^{222}Rn}, \ce{^{218}Po}, \ce{^{214}Po} and \ce{^{210}Po} activity behaviors have been observed thanks to $\alpha$ measurements, putting the acrylic samples in front of the silicon detector. Gamma measurements of the \ce{^{210}Pb} \SI{46.5}{keV} characteristic line ($4\%$ branching ratio) have been also performed by means of the HPGe detector  to quantify the \ce{^{210}Pb} fraction that gravitationally deposited on the samples surface and the deeper implanted one due to its parents decays. The performed measurement sequence is identical for all the studied acrylic samples. After $3$ minutes from the extraction of a sample, $11$ fast alpha measurements in a row (from $3$ to $40$ minutes) have been performed to study \ce{^{218}Po} and \ce{^{214}Po} activity evolution. These measurements have been followed by a long (from $13$ to $42$ days) $\alpha$ measurement to study \ce{^{210}Po} activity evolution in time. In Table~\ref{tab:2}, the sequence of the fast $\alpha$ measurements is reported; in Table~\ref{tab:3}, the duration of the long $\alpha$ measurement of each acrylic sample is shown. Subsequently, two $\gamma$ measurements of the \ce{^{210}Pb} \SI{46.5}{keV} line have been carried out, the second after a rinsing of the plate with distilled water, in order to remove the dust laying on its surface. The surface cleaning allowed to estimate the fraction of \ce{^{210}Pb} that implanted in depth inside the samples.

\begin{table}
\caption{Arrangement of the $\alpha$ measurements for each acrylic sample.}
\label{tab:2} 
\centering
\begin{tabular}{ccc}
\hline\noalign{\smallskip}
& Duration & Observed Isotopes\\
\noalign{\smallskip}.
& [min] \\
\noalign{\smallskip}\hline\noalign{\smallskip}
Meas. 1 (M1)& 3 & \ce{^{218}Po}, \ce{^{214}Po}\\
Meas. 2 (M2)& 3 & \ce{^{218}Po}, \ce{^{214}Po}\\
Meas. 3 (M3)& 3 & \ce{^{218}Po}, \ce{^{214}Po}\\
Meas. 4 (M4)& 10 & \ce{^{214}Po}\\
Meas. 5 (M5)& 10 & \ce{^{214}Po}\\
Meas. 6 (M6)& 20 & \ce{^{214}Po}\\
Meas. 7 (M7)& 20 & \ce{^{214}Po}\\
Meas. 8 (M8)& 20 & \ce{^{214}Po}\\
Meas. 9 (M9)& 40 & \ce{^{214}Po}\\
Meas. 10 (M10)& 40 & \ce{^{214}Po}\\
Meas. 11 (M11)& 40 & \ce{^{214}Po} \\
Long Meas. (LM) & see Table~\ref{tab:3} & \ce{^{210}Po} \\
\noalign{\smallskip}\hline
\end{tabular}
\end{table}

\begin{table}
\caption{Duration of the long $\alpha$ measurements (LM) for each acrylic sample.}
\label{tab:3} 
\centering
\begin{tabular}{ccc}
\hline\noalign{\smallskip}
Sample & Duration & Observed Isotopes\\
\noalign{\smallskip}
 & [days] \\
\noalign{\smallskip}\hline\noalign{\smallskip}
Smooth & 26 & \ce{^{210}Po}\\
Opaque 1& 13 & \ce{^{210}Po}\\
Opaque 2& 21& \ce{^{210}Po}\\
Opaque 3&  21& \ce{^{210}Po}\\
\noalign{\smallskip}\hline
\end{tabular}
\end{table}

\section{Analysis of Acquired Data}
\label{sec:3}

When a sample is extracted from the Rn-Box, it is no longer exposed to the rich \ce{^{222}Rn} atmosphere on the inside. Thus, the secular equilibrium condition on the plate surface is not verified anymore. It is expected that the extracted sample is contaminated by \ce{^{222}Rn} and its daughters, which start to decay following their characteristic half-life. In the following sections, all the analysis passages to study \ce{^{222}Rn} contamination are introduced and discussed.

\subsection{Study of \ce{^{218}Po} and \ce{^{214}Po} Contamination}
\label{Po218-Po214}

\begin{figure}
\includegraphics[scale = 0.45]{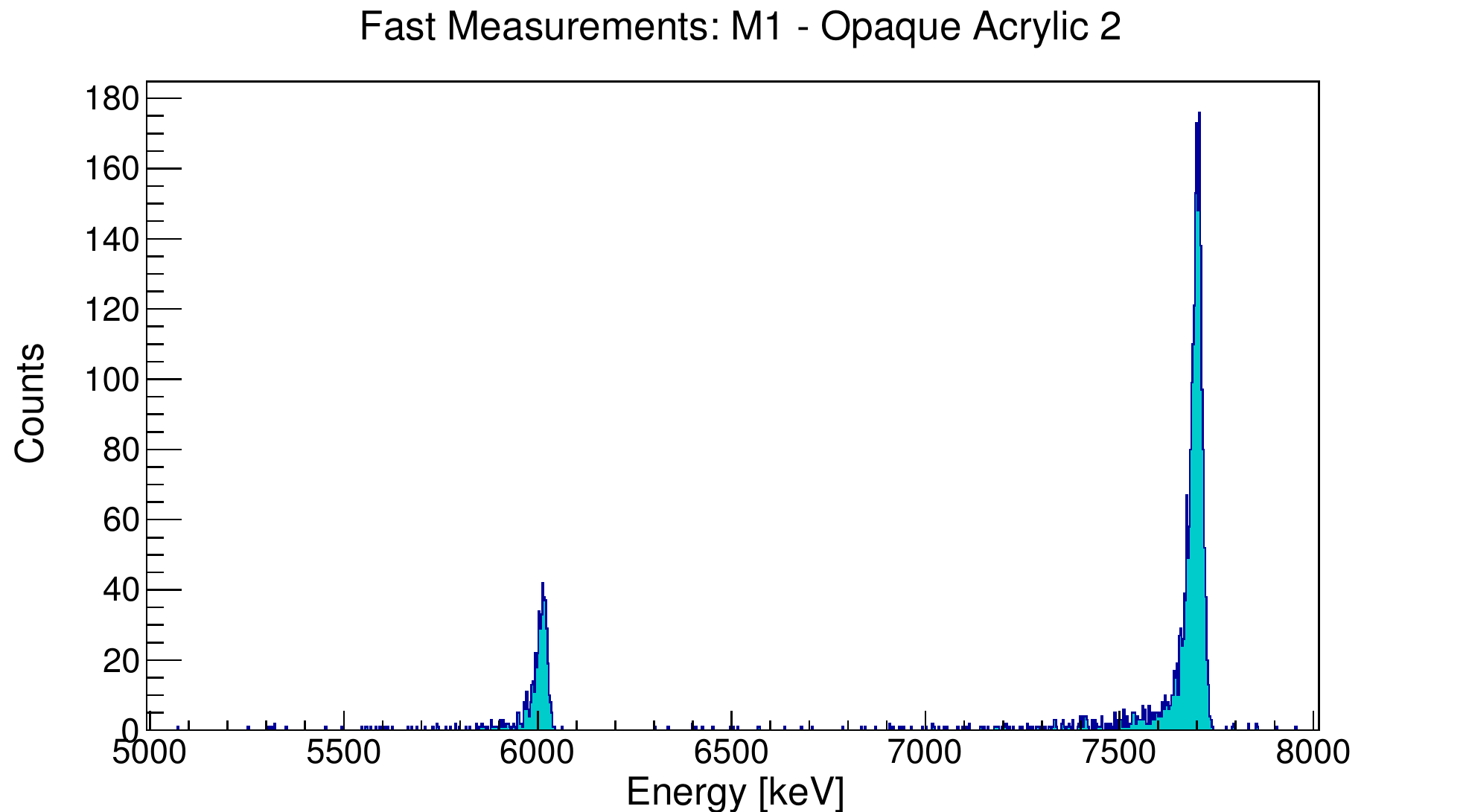}
\caption{Example of an acquired spectrum (M1). \ce{^{218}Po} and \ce{^{214}Po} peaks are visible at~$6.0$\,\si{MeV} and~$7.7$\,\si{MeV}, respectively.}
\label{fig:2}
\end{figure}

The fast $\alpha$ measurements (see Table~\ref{tab:2}) allowed to reconstruct the evolution in time of \ce{^{218}Po} and \ce{^{214}Po} activity. In Figure~\ref{fig:2}, the first acquired spectrum (M1) of one of the analyzed samples is shown, where \ce{^{218}Po} and \ce{^{214}Po} peaks at \SI{6.0}{MeV} and \SI{7.7}{MeV}, respectively, are clearly recognizable. Conversely, there is no evidence of the \ce{^{222}Rn} signal at \SI{5.5}{MeV}. As detailed later in this Section, these measurements allowed to understand the role of the decay type in the contamination process. 

\begin{figure}
\includegraphics[scale = 0.45]{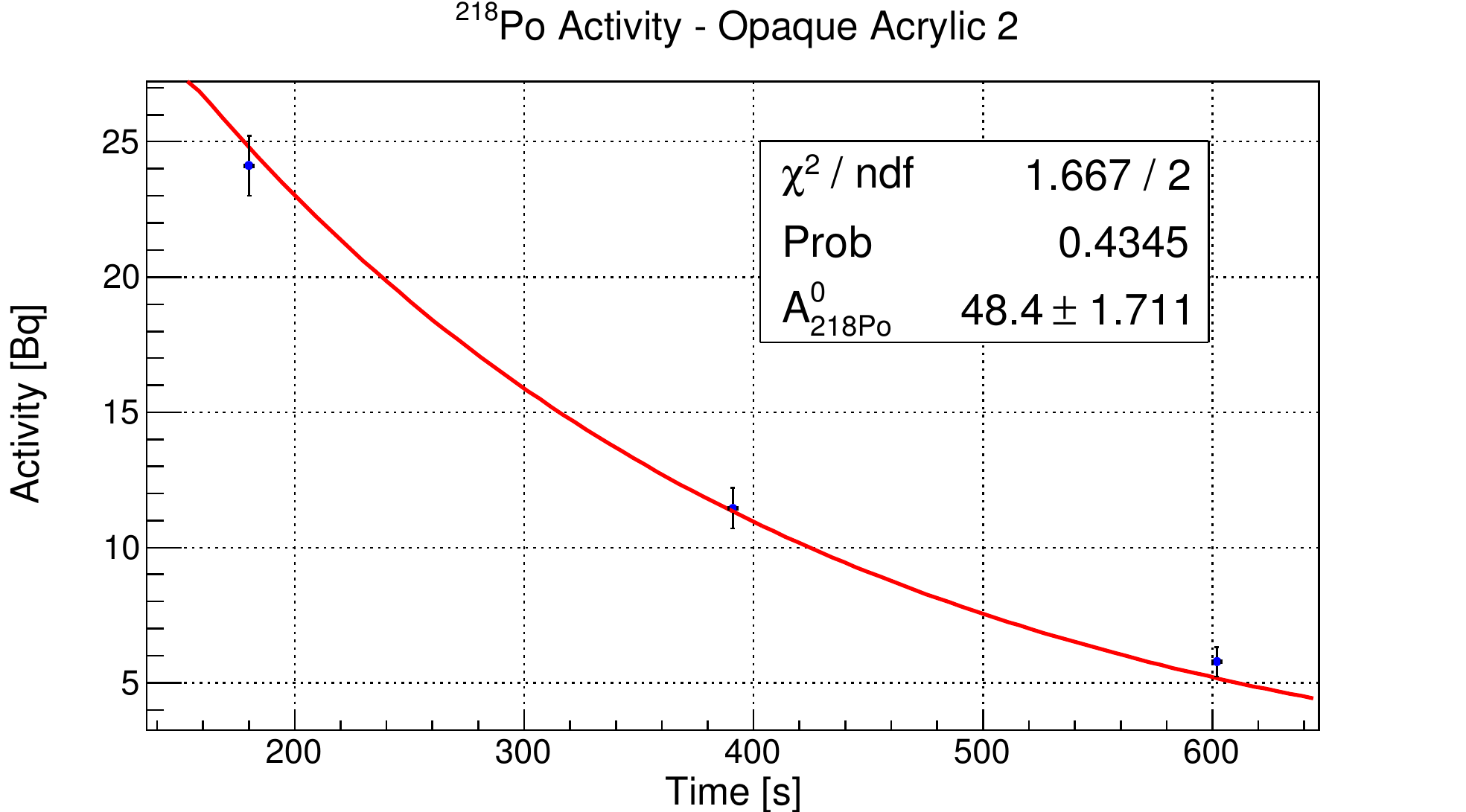}
\caption{Fit of reconstructed \ce{^{218}Po} activity at the beginning of M1, M2 and M3 for the Opaque Acrylic 2, with the $A^{0}_{218Po}$ estimation.}
\label{fig:3}
\end{figure}

By the first three $\alpha$ measurements (M1, M2 and M3), it has been possible to extrapolate \ce{^{218}Po} activity value on each sample at the moment of extraction ($A^{0}_{218Po}$). The instantaneous activities obtained from M1, M2 and M3 spectra have been successfully interpolated with a decreasing exponential function. In Figure~\ref{fig:3}, the interpolation of Opaque Acrylic 2 activities is shown as an example. The absence of both a growing term of \ce{^{218}Po} and a recognizable \ce{^{222}Rn} peak suggests that radon diffusion was not a competing contamination process in this case. However, more on \ce{^{222}Rn} diffusion will be discussed later in Section~\ref{RnDiff}.

For each sample, the reconstructed $A^{0}_{218Po}$ values are reported in Table~\ref{tab:4}. The errors on $A^{0}_{218Po}$ are those extrapolated from the fit procedure (stat) together with a systematic uncertainty (syst) coming from the observed fluctuations of \ce{^{222}Rn} concentration in the chamber (see Fig.~\ref{fig:RnMon}). The systematic uncertainty has been chosen as half of the distance between maximum and minimum detected values and it is equal to $5.4\%$. Since the $A^{0}_{218Po}$ values are consistent, the initial conditions of all the samples can be considered homogeneous. As a consequence, the contamination process can be considered replicable, allowing the comparison of the results obtained for the following \ce{^{222}Rn} daughters in the chain.

\begin{table}
\caption{$A^{0}_{218Po}$ estimation for each measured acrylic sample.}
\label{tab:4}  
\centering
\begin{tabular}{cc}
\hline\noalign{\smallskip}
Sample & $A^{0}_{218Po}$ ($\pm$ stat $\pm$ syst)\\
\noalign{\smallskip}
 & [Bq] \\
\noalign{\smallskip}\hline\noalign{\smallskip}
Smooth & 46.8 $\pm$ 1.6 $\pm$ 2.5 \\
Opaque 1& 37.2 $\pm$ 1.5 $\pm$ 2.0 \\
Opaque 2& 48.4 $\pm$ 1.7 $\pm$ 2.6 \\
Opaque 3& 42.9 $\pm$ 1.6 $\pm$ 2.3 \\
\noalign{\smallskip}\hline
\end{tabular}
\end{table}

\begin{figure}
\includegraphics[scale =0.45]{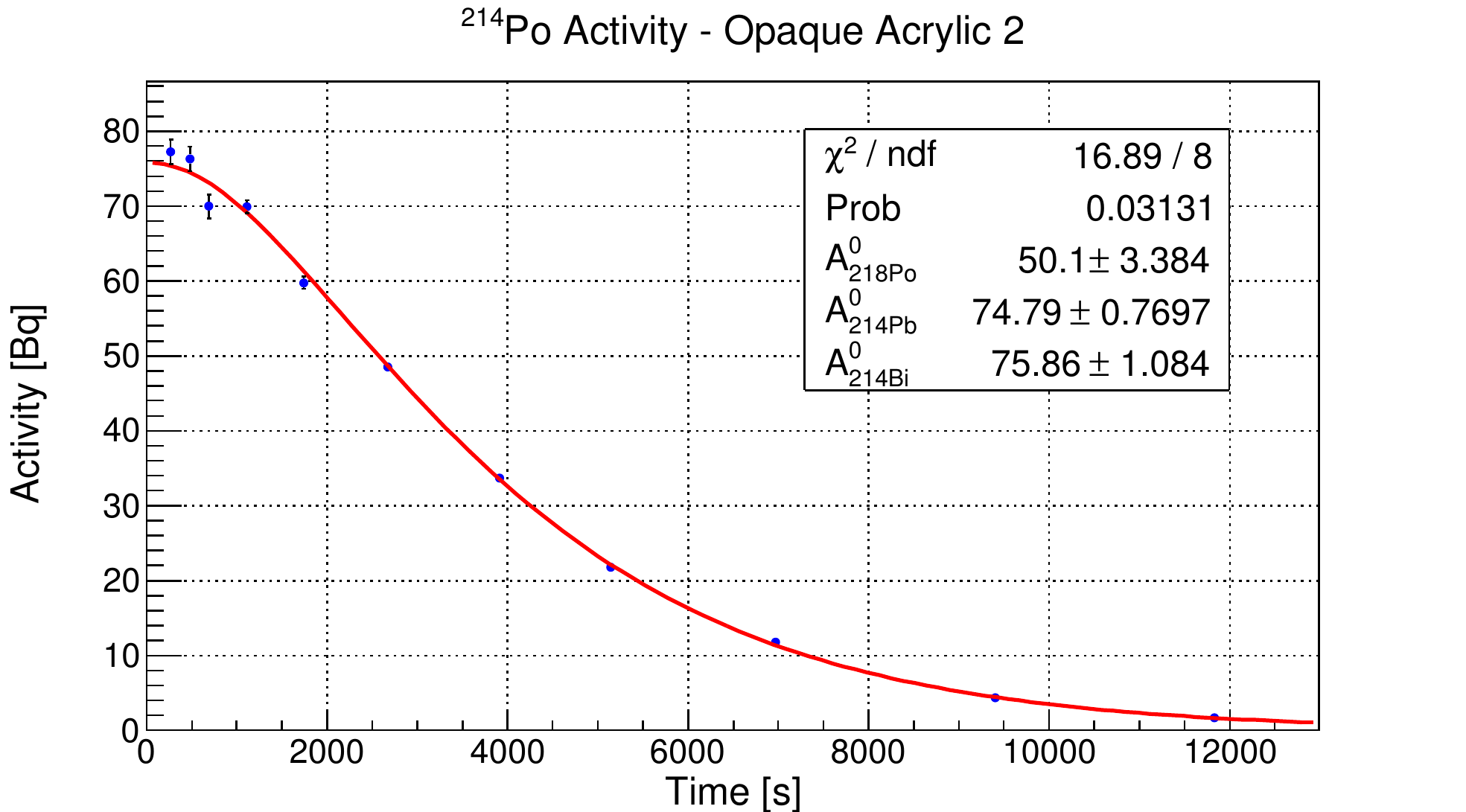}
\caption{Fitted plot of the \ce{^{214}Po} mean activities for the Opaque Acrylic 2 sample. The goodness of fit is acceptable, and $A^{0}_{214Pb}$, $A^{0}_{214Bi}$ result to be compatible.}
\label{fig:4}
\end{figure}

\ce{^{214}Po} mean activity has been also obtained for each of the $11$ fast $\alpha$ measurements. The activity values have been fitted using Equation~\eqref{Bateman}, summing three independent terms: One chain originating from \ce{^{218}Po}, present on the acrylic at the moment of extraction; one chain starting from \ce{^{214}Pb}; and the \ce{^{214}Bi} contribution. In Figure~\ref{fig:4}, an example of interpolation is reported. The activity values at the moment of extraction ($A^{0}_{218Po}$, $A^{0}_{214Pb}$ and $A^{0}_{214Bi}$) have been extrapolated for each sample; $A^{0}_{214Pb}$ and $A^{0}_{214Bi}$ have been left as free parameters, while $A^{0}_{218Po}$ has been forced in one standard deviation range around the value extrapolated from the fit of \ce{^{218}Po} measured activities. $A^{0}_{214Pb}$ and $A^{0}_{214Bi}$ obtained values are reported in Table~\ref{tab:5}: they turn out to be compatible in each measured sample.
 
\begin{table}
\caption{$A^{0}_{214Pb}$ and $A^{0}_{214Bi}$ estimations from \ce{^{214}Po} activity fit for each measured acrylic sample.}
\label{tab:5} 
\centering
\begin{tabular}{ccc}
\hline\noalign{\smallskip}
Sample & $A^{0}_{214Pb}$ & $A^{0}_{214Bi}$ \\
\noalign{\smallskip}
& [Bq] & [Bq] \\
\noalign{\smallskip}\hline\noalign{\smallskip}
Smooth & 79.8 $\pm$ 0.8 & 80.0 $\pm$ 1.0\\
Opaque 1& 54.3 $\pm$ 0.7 & 55.7 $\pm$ 0.9 \\
Opaque 2& 74.8 $\pm$ 0.8 & 76.0 $\pm$ 1.0\\
Opaque 3& 65.7 $\pm$ 0.7 & 68.0 $\pm$ 1.0\\
\noalign{\smallskip}\hline
\end{tabular}
\end{table}

The outcomes of the \ce{^{218}Po} and \ce{^{214}Po} measurements highlight the dependency of the contamination mechanism on the decay type. Since \ce{^{218}Po} activity is always lower than \ce{^{214}Pb} activity, both the $\alpha$ decays of \ce{^{222}Rn} and \ce{^{218}Po} must be responsible for the implantation of the recoiling nuclei inside the acrylic. On the other hand, since \ce{^{214}Pb} and \ce{^{214}Bi} activities are compatible, the implantation is not affected by the $\beta$ decays of these isotopes. Thus, the key outcome of these measurements is that the only decay implanting isotopes inside the acrylic is the alpha one, and subsequent alpha decays add their contribution to the measured total activity of the contaminant. Conversely, the beta decays do not cause any implantation of radioactive contaminants.

The third $\alpha$-emitter in the \ce{^{222}Rn} chain is \ce{^{214}Po}. Given its short half-life ($\tau_{1/2} = \SI{163}{\micro s}$), its contribution to the sample contamination can be inferred only by studying its parent (\ce{^{214}Bi}) and daughter (\ce{^{210}Pb}) nuclides. In the \ce{^{222}Rn} chain evolution, \ce{^{210}Pb} can be considered as an almost stable element that accumulates in the samples during the whole exposure period in the Rn-Box. The activity of \ce{^{210}Pb} has been directly quantified by means of the $\gamma$ measurements described in Section~\ref{sec:Pb210-contamination}, whose results are recalled for convenience in Table~\ref{tab:6} as MA (Measured Activity); here, the errors also include a $5\%$ systematic uncertainty due to the Monte Carlo simulation for the efficiency evaluation, besides the statistical contribution given in Table~\ref{tab:8}. On the other hand, \ce{^{214}Bi} activity in each sample can also be used to predict an expected \ce{^{210}Pb} count rate, which does not take into account the additional contribution of \ce{^{214}Po} direct implantation. Therefore, a comparison between the measured and expected \ce{^{210}Pb} activities should provide a hint for the eventual \ce{^{214}Po} implantation. 
The expected activity of \ce{^{210}Pb} in a sample can be calculated as:
\begin{equation}
\label{eq:Pb210-prediction}
A_{210Pb}^{0} = \lambda_{210Pb} A_{214Bi}^{0} \Delta t_{exp} 
\end{equation}
where $\Delta t_{exp}$ is the exposure period of the single sample and $A_{214Bi}^{0}$ is \ce{^{214}Bi} activity estimation of Table~\ref{tab:5}. In Equation~\eqref{eq:Pb210-prediction}, it is assumed that secular equilibrium between \ce{^{214}Bi} and \ce{^{214}Po} (i.e., $A_{214Bi}^{0} = A_{214Po}^{0}$) is instantaneously reached, a condition which always holds because of the short \ce{^{214}Po} half-life. \ce{^{210}Pb} activities predicted by Equation~\eqref{eq:Pb210-prediction} are also reported in Table~\ref{tab:6} as RA (Reconstructed Activity). Unfortunately, as evident from the MA and RA values, a conclusive statement about \ce{^{214}Po} recoiling nuclei implantation on the acrylic samples is not possible. The role of the dust, to which electrically charged radon daughters stick, as well as the cleaning procedure performed to remove it from the samples surfaces (cf. Sec.~\ref{sec:Pb210-contamination} and Table~\ref{tab:8}) avoid the disentangling of the independent \ce{^{214}Po} implantation mechanism and the quantification of its effect.

\begin{table}
\caption{\ce{^{210}Pb} measured activities (MA) and reconstructed activities (RA). MAs have been obtained after cleaning the sample surface (see Sec.~\ref{sec:Pb210-contamination}).}
\label{tab:6} 
\centering
\begin{tabular}{ccc}
\hline\noalign{\smallskip}
Sample & \ce{^{210}Pb} MA &  \ce{^{210}Pb} RA \\
\noalign{\smallskip}
& [Bq] & [Bq] \\
\noalign{\smallskip}\hline\noalign{\smallskip}
Smooth & 0.40 $\pm$ 0.03 & 0.59 $\pm$ 0.04 \\
Opaque 1& 0.49 $\pm$ 0.04 & 0.34 $\pm$ 0.02 \\
Opaque 2& 0.56 $\pm$ 0.03 & 0.58 $\pm$ 0.04 \\
Opaque 3& 0.25 $\pm$ 0.03 & 0.23 $\pm$ 0.02 \\
\noalign{\smallskip}\hline
\end{tabular}
\end{table}

\subsection{Study of \ce{^{210}Po} Contamination}
\label{sec:Po210-implantation}

\begin{figure}
\includegraphics[scale = 0.45]{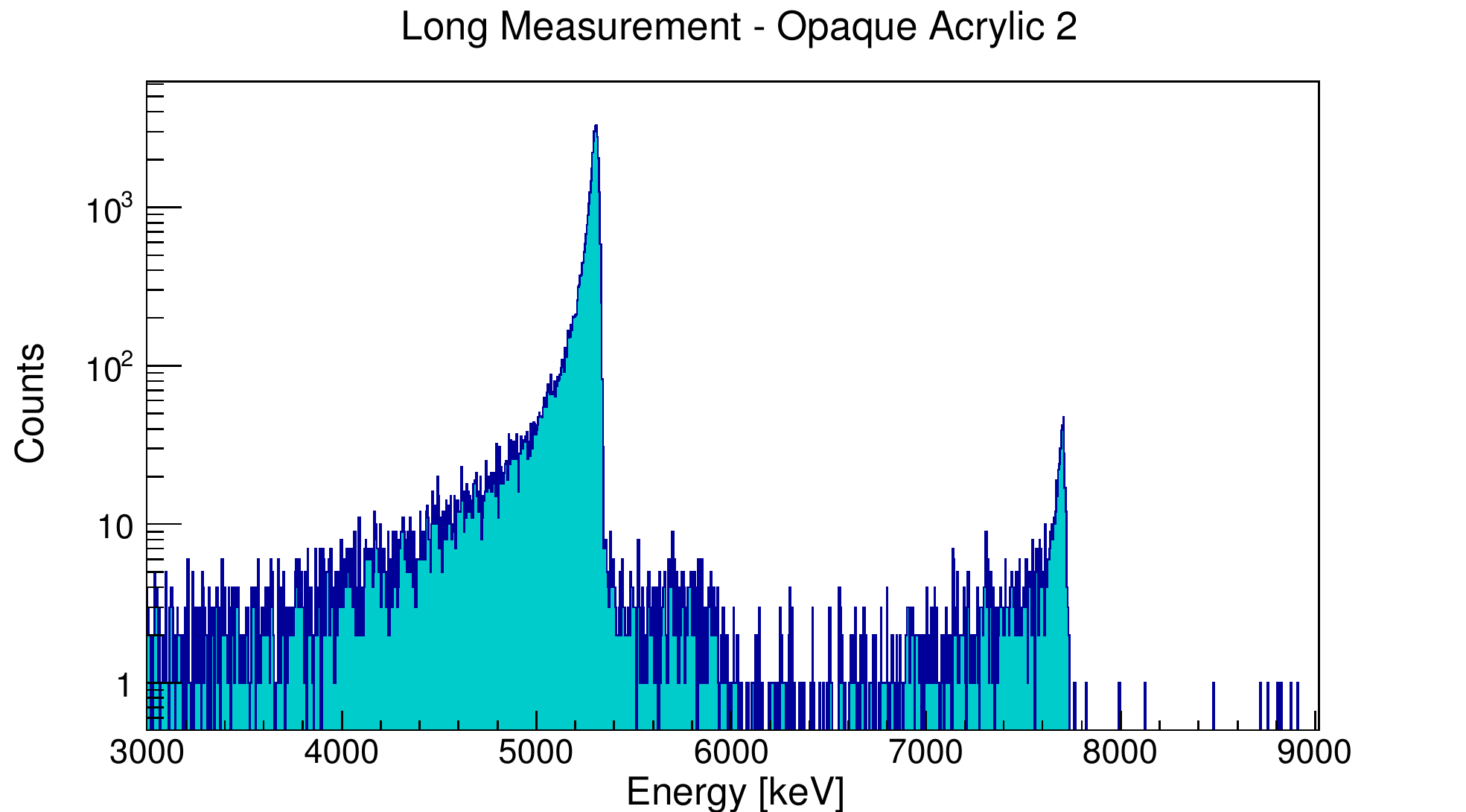}
\caption{Acquired spectrum of a 21 days long $\alpha$ measurement. At \SI{5.3}{MeV} the \ce{^{210}Po} peak is well visible, with a long tail at lower energies. At about \SI{7.7}{MeV} the \ce{^{214}Po} peak is still recognizable.}
\label{fig:5}
\end{figure}

The several days long $\alpha$ measurements (see Table~\ref{tab:3}) allowed a study of the evolution in time of \ce{^{210}Po} activity. This isotope decays via $\alpha$ channel with characteristic energy of \SI{5.3}{MeV}. In Figure~\ref{fig:5}, an example of an acquired spectrum in the~$3.0$-$9.0$~\si{MeV} region is shown.

\begin{figure}
\includegraphics[scale = 0.45]{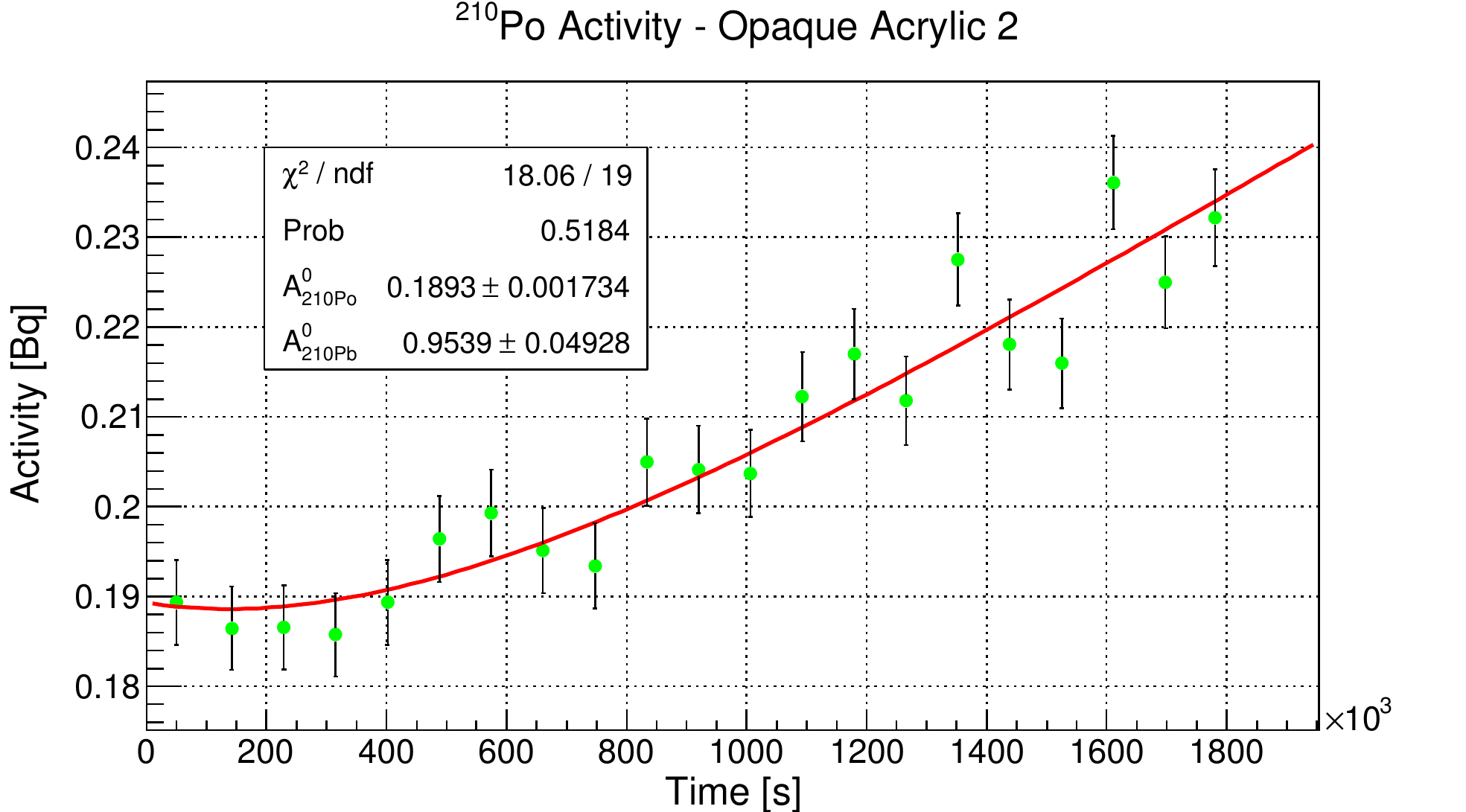}
\caption{Fit of the \ce{^{210}Po} activity plot for Opaque Acrylic 2 (before cleaning). The goodness of fit is acceptable, and the $A^{0}_{210Po}$ and $A^{0}_{210Pb}$ activity values at the time $t = 0$ have been extrapolated.}
\label{fig:6}
\end{figure}

The measurable activity of \ce{^{210}Po} is composed of two different contributions: The first one comes from the \ce{^{210}Po} originated inside the chamber which contaminated the acrylic directly, and follows the exponential law characterized by its own $\tau_{1/2}$; the second one is generated by the decay of \ce{^{210}Pb} inside the sample, and increases in time towards the condition $A_{210Po} = A_{210Pb}$. In fact, once a sample is extracted from the Rn-Box, the \ce{^{210}Pb} present in the plate starts to decay following its half-life, generating its own chain (accordingly with Eq.~\eqref{Bateman}). In order to verify this behavior, one day long subsequent $\alpha$ measurements have been exploited to track the growth of \ce{^{210}Po} activity towards the \ce{^{210}Pb} one. The daily mean activity values have been interpolated leaving as free parameters \ce{^{210}Po} and \ce{^{210}Pb} activity values at the extraction time (Fig.~\ref{fig:6}). For each sample, the $A^{0}_{210Pb}$ values obtained from the interpolation are consistent with the corresponding \ce{^{210}Pb} activity values extracted from the $\gamma$ measurements (see Sec.~\ref{sec:Pb210-contamination} and~\ref{sec:Po210-cont-clean}).

\subsection{Study of Polonium Isotopes Implantation}
\label{sec:Po-contamination}

As evident by the comparison of Figures~\ref{fig:2} and~\ref{fig:5}, \ce{^{210}Po} signal centered at \SI{5.3}{MeV} shows a more enhanced tail with respect to \ce{^{218}Po} and \ce{^{214}Po}. Since alpha decays are the only responsible of the nuclear implantation inside the acrylic and \ce{^{210}Po} is the last radioactive contaminant in the \ce{^{222}Rn} chain, it is plausible to expect a deeper implantation of this isotope inside the samples. To quantify the penetration of polonium isotopes inside acrylic, the ratio of the tail component to the peak one has been calculated. For \ce{^{210}Po} the \emph{Tail-to-Peak Ratio} (TPR) is defined as the percentage ratio of the spectrum integral in the~$5.0$-$5.2$~\si{MeV} and~$5.2$-$5.4$~\si{MeV} energy regions, respectively: 
\begin{equation}
\label{eq:1}
\text{TPR}_{210Po} = \frac{\text{No. of Counts}~(5.0\text{-}5.2~\si{MeV})}{\text{No. of Counts}~(5.2\text{-}5.4~\si{MeV})} \cdot 100
\end{equation}
Similarly, for \ce{^{218}Po} and \ce{^{214}Po} TPR has been defined as follows: 
\begin{equation}
\text{TPR}_{218Po} = \frac{\text{No. of Counts}~(5.7\text{-}5.9~\si{MeV})}{\text{No. of Counts}~(5.9\text{-}6.1~\si{MeV})} \cdot 100
\end{equation}
and
\begin{equation}
\text{TPR}_{214Po} = \frac{\text{No. of Counts}~(7.4\text{-}7.6~\si{MeV})}{\text{No. of Counts}~(7.6\text{-}7.8~\si{MeV})} \cdot 100
\end{equation}
For each isotope, the TPR has been calculated for each acquired spectrum (from M1 to M11 for \ce{^{218}Po} and \ce{^{214}Po}, and each daily saving for \ce{^{210}Po}) and its trend in time has been verified to be constant; thus, both the tail and the peak component of polonium signals grow with the same rate. As a consequence, during the measurement period no mechanism interfered with the contamination features, changing the reciprocal radioactivity amounts on the sample. In Table~\ref{tab:7}, $\text{TPR}_{210Po}$, $\text{TPR}_{214Po}$ and $\text{TPR}_{218Po}$ mean values of all the samples are reported.

\begin{table}
\caption{$\text{TPR}_{210Po}$, $\text{TPR}_{218Po}$ and $\text{TPR}_{214Po}$ estimations for each acrylic sample.}
\label{tab:7} 
\centering
\begin{tabular}{cccc}
\hline\noalign{\smallskip}
Sample & $\text{TPR}_{210Po}$ & $\text{TPR}_{214Po}$ & $\text{TPR}_{218Po}$ \\
\noalign{\smallskip}
& [\%] & [\%] & [\%] \\
\noalign{\smallskip}\hline\noalign{\smallskip}
Smooth & 3.9 $\pm$ 0.1 &  3.4 $\pm$ 0.1 & 2.0 $\pm$ 0.5\\
Opaque 1& 13.7 $\pm$ 0.4  & 7.9 $\pm$ 0.2 & 6.6 $\pm$ 1.1 \\
Opaque 2& 11.4 $\pm$ 0.2 & 6.7 $\pm$ 0.1 & 7.0 $\pm$ 1.0 \\
Opaque 3& 14.3 $\pm$ 0.5 & 8.4 $\pm$ 0.2 & 7.0 $\pm$ 1.0 \\
\noalign{\smallskip}\hline
\end{tabular}
\end{table}

The TPR calculation confirms the deeper implantation of \ce{^{210}Po} inside the acrylic, due to three subsequent $\alpha$ decays. $\text{TPR}_{210Po}$ values are in fact clearly higher than $\text{TPR}_{214Po}$ and $\text{TPR}_{218Po}$. Moreover, there is a substantial difference between the results of the smooth plate and the opaque ones, that denotes an important dependence of the implantation on the surface texture. In fact, on an opaque surface even a superficial contamination is located at different depths, because of the peak-valley structure of its irregular texture. On the other hand, a smooth surface does not provide this augmented depth, resulting in a more superficial contamination.

\subsection{Study of \ce{^{210}Pb} Contamination}
\label{sec:Pb210-contamination}

Lead-$210$, being the longest living isotope of \ce{^{222}Rn} chain ($\tau_{1/2} = \SI{22.3}{y}$), will stay within the sample for a very long time. A deep contamination cannot be generally removed from the sample and the only way to get rid of this component is waiting for the complete decay of the implanted radiocontaminants. For this reason, a determination of the \ce{^{210}Pb} fraction deeply implanted in acrylic is fundamental. This was achieved by comparing the experimental \ce{^{210}Pb} activity~---~inferred from the \SI{46.5}{keV} $\gamma$-line rate~---~of each sample as extracted from the Rn-Box and after the surface cleaning. For the surface cleaning, each acrylic plate has been rinsed with distilled water for about two minutes, thus removing the \ce{^{210}Pb} component deposited on the plates together with dust. After that, a spectroscopic gamma measurement has been performed on each sample. In Table~\ref{tab:8}, \ce{^{210}Pb} activities are presented before and after the dust removal, together with the estimated percentage of the deep component with respect to the total one. Since the \emph{Deep Fraction} (D.F.) is about $67\%$, the majority of the \ce{^{210}Pb} contamination comes from a deep implantation of this isotope inside the acrylic that cannot be removed by a simple cleaning.

\begin{table}
\caption{\ce{^{210}Pb} activities estimated before (BC) and after (AC) the cleaning of each sample surface, with the Deep Fraction (D.F.) component.}
\label{tab:8}
\centering
\begin{tabular}{cccc}
\hline\noalign{\smallskip}
Sample & BC Activity & AC Activity & D.F.\\
\noalign{\smallskip}
& [Bq] & [Bq] & [\%] \\
\noalign{\smallskip}\hline\noalign{\smallskip}
Smooth & 0.61 $\pm$ 0.03 &  0.40 $\pm$ 0.02 & 65 $\pm$ 5\\
Opaque 1& 0.68 $\pm$ 0.04  & 0.49 $\pm$ 0.03 & 73 $\pm$ 5\\
Opaque 2& 0.89 $\pm$ 0.05 & 0.56 $\pm$ 0.03 & 63 $\pm$ 5\\
Opaque 3& 0.38 $\pm$ 0.02 & 0.25 $\pm$ 0.02 & 66 $\pm$ 6\\
\noalign{\smallskip}\hline
\end{tabular}
\end{table}

\begin{figure}
\includegraphics[scale = 0.45]{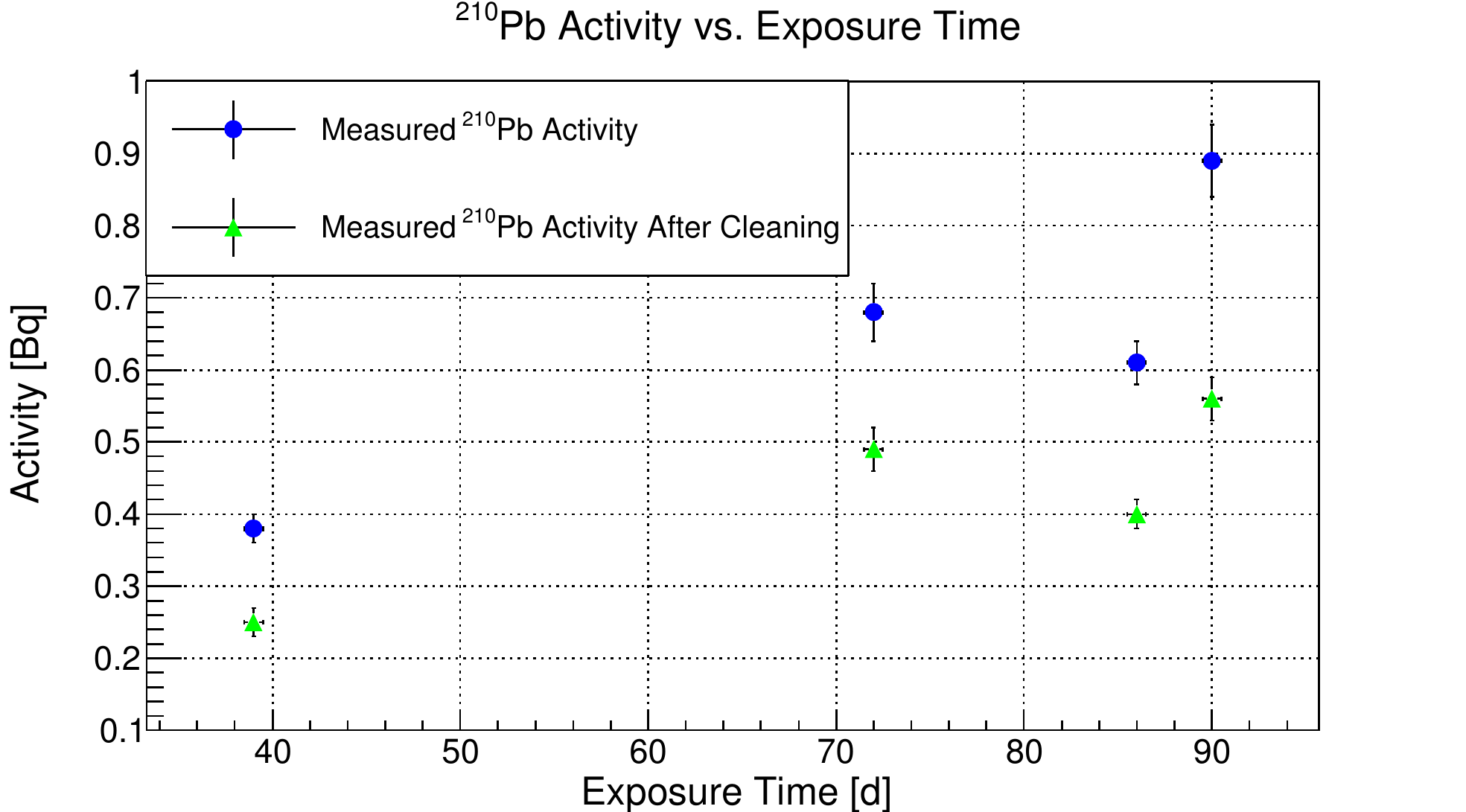}
\caption{\ce{^{210}Pb} measured activities vs. the exposure period for all the samples. The activities before and after cleaning grow with time. The couple of points at $86$ days of exposure is related to the smooth sample which shows lower activity, most likely due to the smaller effective area.}
\label{fig:7}
\end{figure}

\begin{figure}
\includegraphics[scale = 0.45]{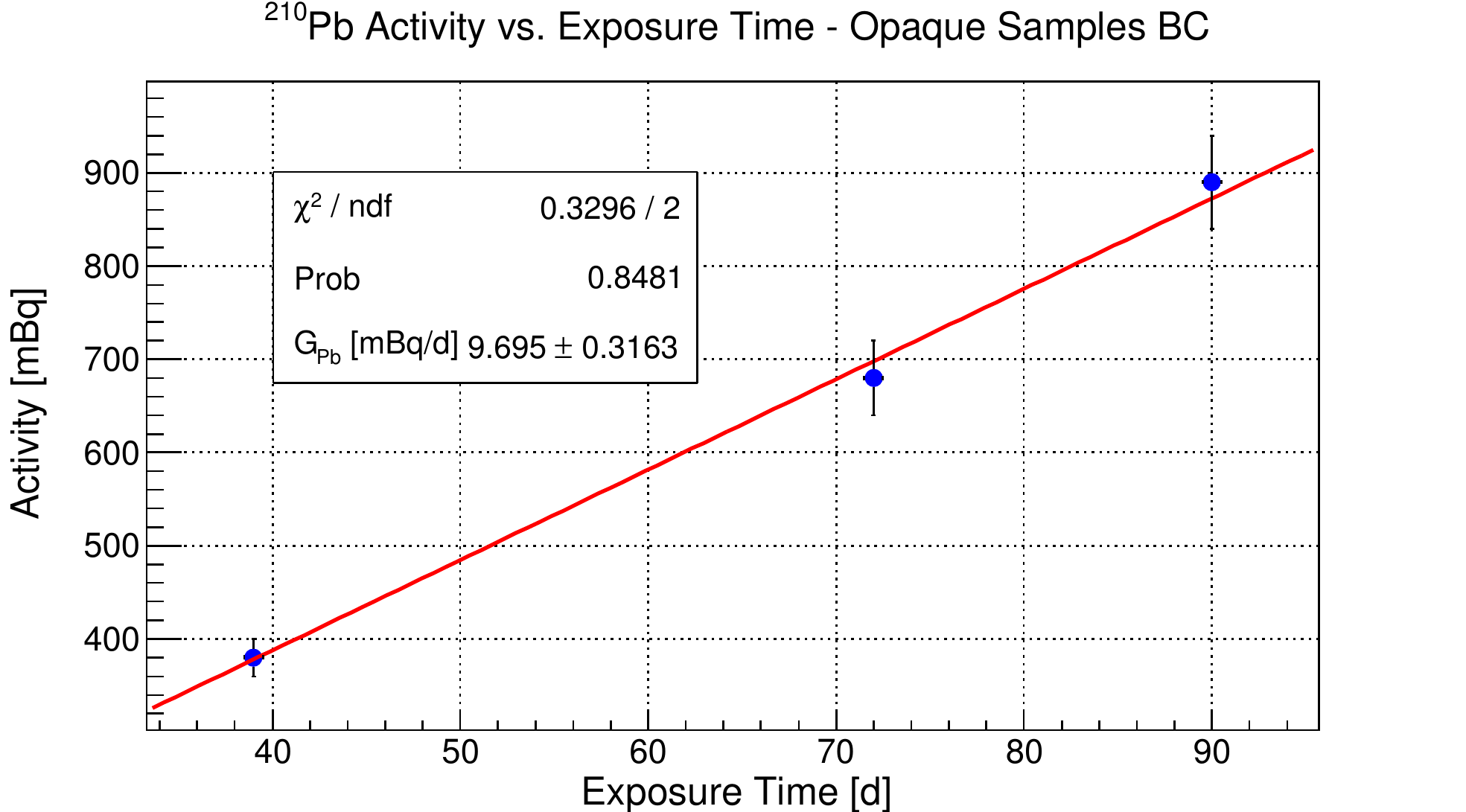}
\caption{Interpolation of \ce{^{210}Pb} activities measured on opaque samples before cleaning (BC) their surfaces.}
\label{fig:FitPb_BC}
\end{figure}

\begin{figure}
\includegraphics[scale = 0.45]{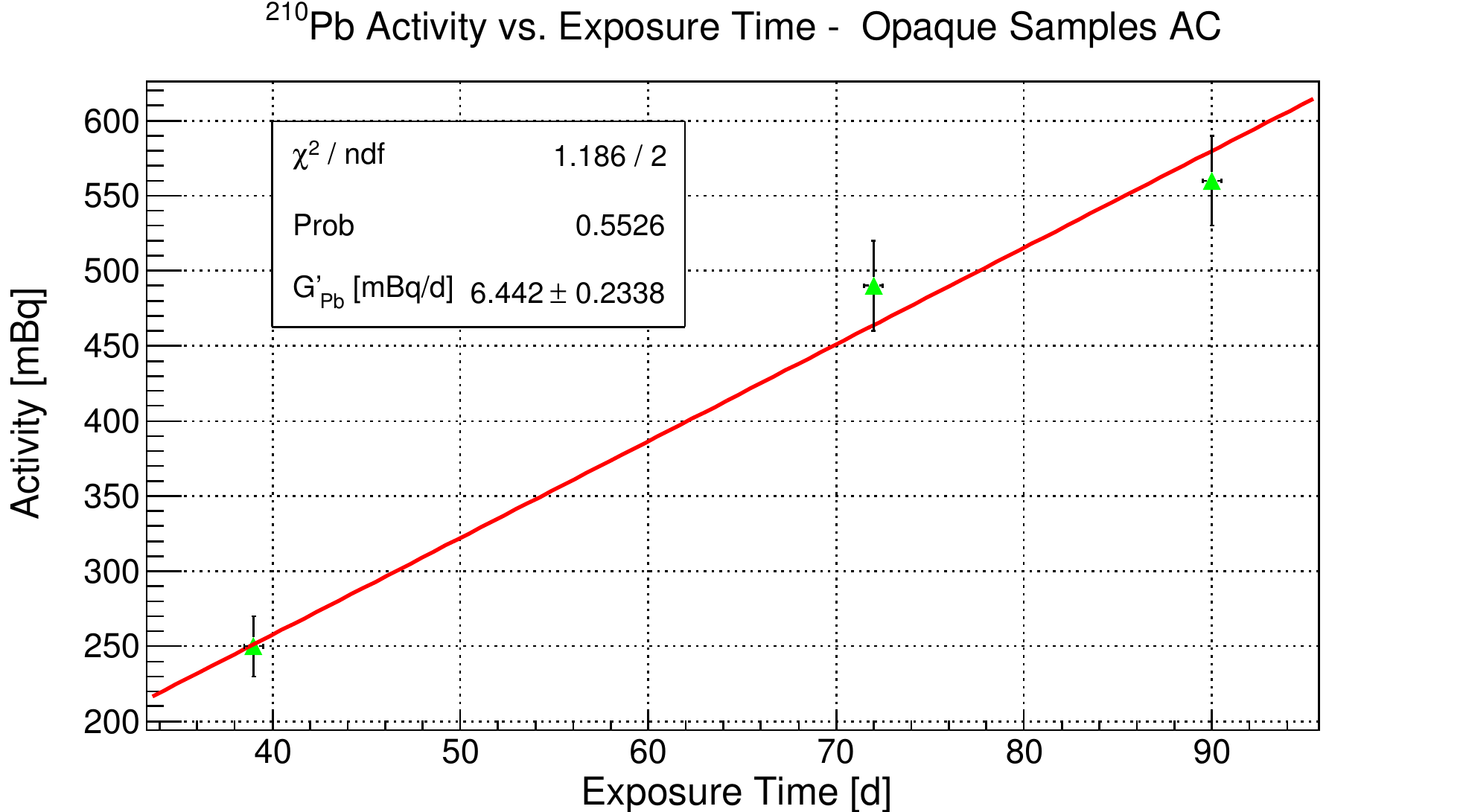}
\caption{Interpolation of \ce{^{210}Pb} activities measured on opaque samples after cleaning (AC) their surfaces.}
\label{fig:FitPb_AC}
\end{figure}

As previously inferred, \ce{^{210}Pb} can be considered as a stable element that accumulates on the samples. Therefore, it can be assumed that the measured \ce{^{210}Pb} activity increases with the time the samples spend inside the chamber, where they are exposed to \ce{^{222}Rn}. In Figure~\ref{fig:7}, a plot of \ce{^{210}Pb} activities before (blue dots) and after (green dots) the cleaning of each sample as a function of the exposure time is shown. The third pair of green and blue dots are \ce{^{210}Pb} activity values before and after the cleaning of the smooth sample and does not follow the trend of the other points. The points related to the opaque samples have been interpolated with a linear function, with the hypothesis that at time of Rn-Box closing there would have been a \ce{^{210}Pb} activity equal to~\SI{0}{Bq}. In Figures~\ref{fig:FitPb_BC} and~\ref{fig:FitPb_AC}, fits of \ce{^{210}Pb} activities before and after the cleaning of opaque samples are shown, respectively.

The experimental condition related to this analysis is of a high \ce{^{222}Rn} concentration environment that surrounds the acrylic plates. This condition brings to a \ce{^{210}Pb} concentration higher than the one that may be found exposing the samples to atmospheric \ce{^{222}Rn} levels. The linearity of \ce{^{210}Pb} accumulation at high activities implies the same behavior at lower levels. As a result of the measurements exposed in this Section, it becomes possible making previsions about the \ce{^{210}Pb} quantity that would contaminate acrylic at different \ce{^{222}Rn} concentrations.

\subsection{Study of \ce{^{210}Po} Contamination after Cleaning}
\label{sec:Po210-cont-clean}

\begin{figure}
\includegraphics[scale =0.45]{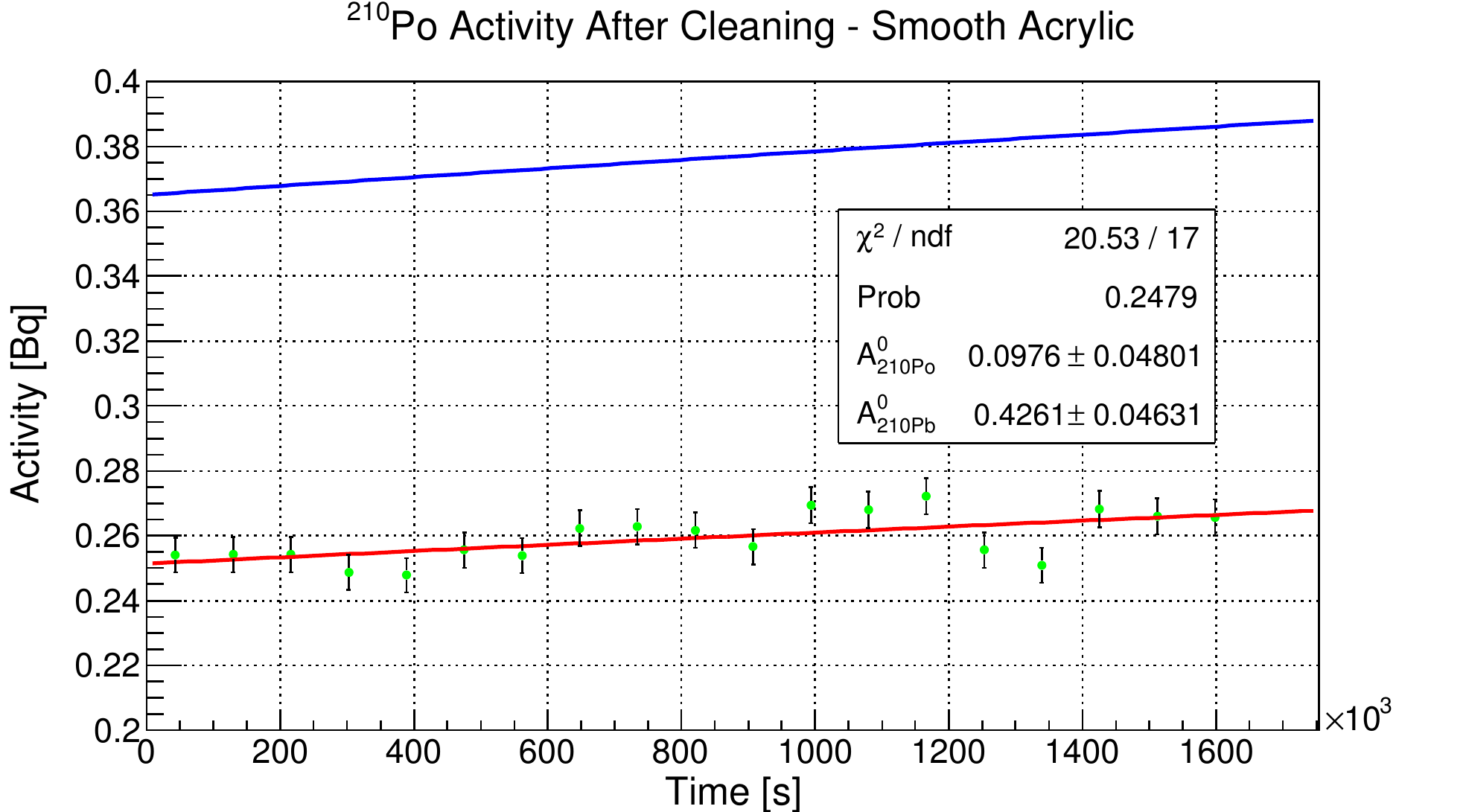}
\caption{Fit of the \ce{^{210}Po} activity after the surface cleaning for the Smooth Acrylic sample (red line), together with the prediction of the \ce{^{210}Po} activity based on fit function of the not cleaned sample. The goodness of the fit is acceptable, and the \ce{^{210}Po} and \ce{^{210}Pb} activity values at the time of the cleaning have been extrapolated ($\tilde{A}^{0}_{210Po}$ and $\tilde{A}^{0}_{210Pb}$, respectively).}
\label{fig:8}
\end{figure}

As a further validation of the obtained results, an additional $\alpha$ measurement after surface cleaning of the smooth sample has been performed. Evolution in time of \ce{^{210}Po} activity has been studied, as before, by exploiting subsequent daily measurements. The daily mean activity points have been interpolated with the same function described in Section~\ref{sec:Po210-implantation} for \ce{^{210}Po} analysis before the cleaning. From the fit (red line in Fig.~\ref{fig:8}) two free parameters ($\tilde{A}^{0}_{210Po}$ and $\tilde{A}^{0}_{210Pb}$) have been extrapolated, representing \ce{^{210}Po} and \ce{^{210}Pb} levels, respectively, immediately after the \ce{H_{2}O} cleaning procedure. The prediction of \ce{^{210}Po} activity evolution without cleaning process has been represented with the blue line in Figure~\ref{fig:8}. The estimated \ce{^{210}Pb} activity value right after the cleaning of the sample ($\tilde{A}^{0}_{210Pb}$) is equal to~$(0.43 \pm 0.05)$\,\si{Bq} and it is compatible with the measured activity of~$(0.40 \pm 0.02)$\,\si{Bq} (see Table~\ref{tab:8}). \ce{^{210}Po} activity estimation ($\tilde{A}^{0}_{210Po}$) right after the sample cleaning turns out to be extremely small, as expected. These results support the consistency of the analysis.

\begin{figure}
\includegraphics[scale =0.45]{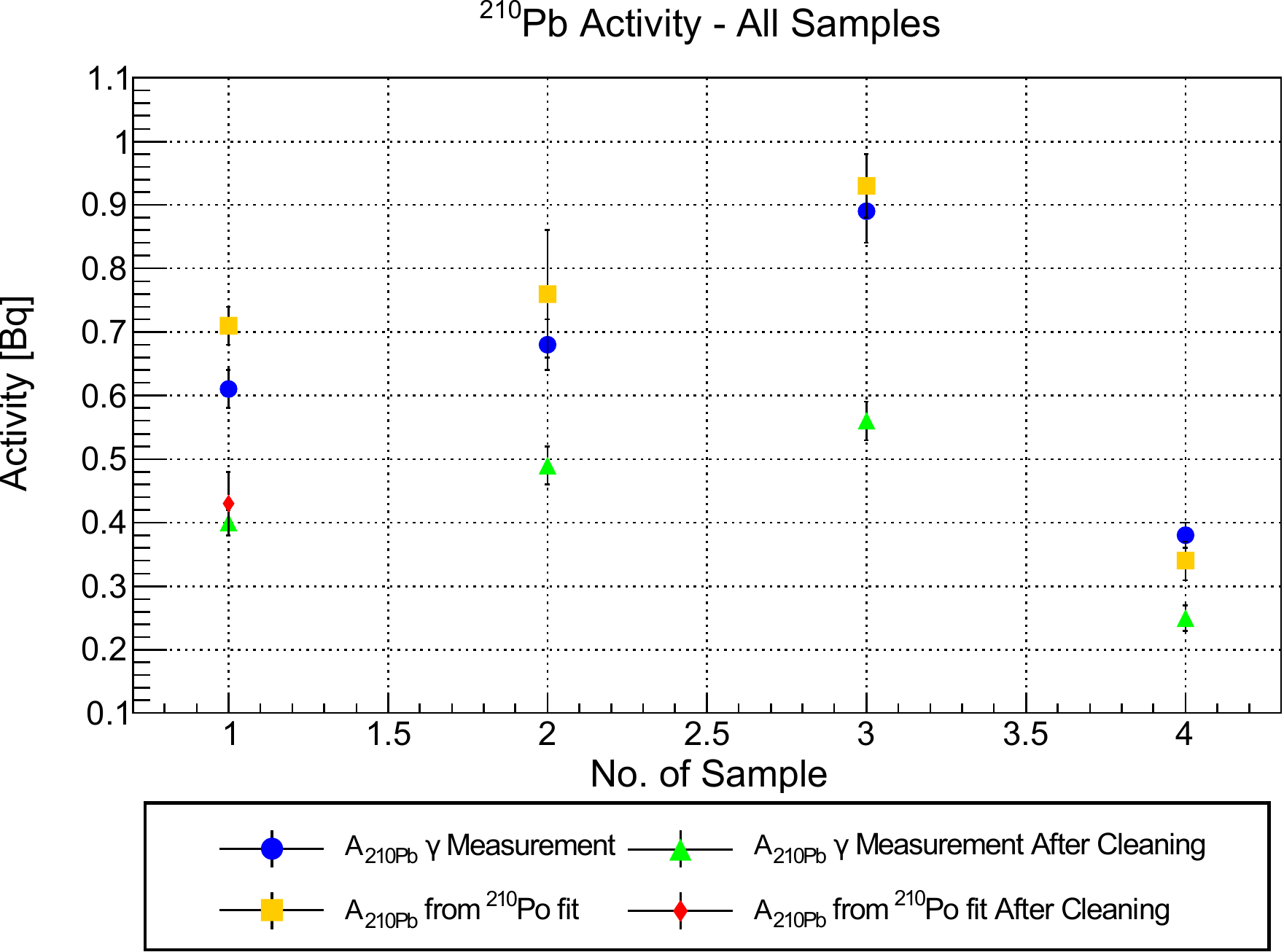}
\caption{Summarizing plot of the measured and valued \ce{^{210}Pb} activities by \ce{^{210}Po} fits before and after the cleaning, for each sample. The measurement of the \ce{^{210}Po} activity after cleaning was performed only for the smooth sample.}
\label{fig:9}
\end{figure}

In Figure~\ref{fig:9}, \ce{^{210}Pb} activities before and after the cleaning extrapolated by \ce{^{210}Po} fits (yellow and red dots, respectively) and measured (blue and green dots, respectively) are reported for each sample. All the measured activity values agree with the previsions obtained by the interpolation methods. The \ce{^{210}Po} activity loss due to the rinse in water is equal to $(31 \pm 3)\%$, calculated as the mean difference between the measured activity points and the corresponding values of the predicted function. This estimation is consistent with the \ce{^{210}Pb} loss effectively measured, equal to $(35 \pm 4)\%$. This result proves that the system has not been modified between the two measurements, and the surface cleaning has been the only cause of activity loss.

In addition, the TPR value of the \ce{^{210}Po} signal after the acrylic surface cleaning ($\text{TPR}_{210Po}^{AC}$) has been calculated following Equation~\eqref{eq:1}. An increase of this value is expected because of the removal of the \ce{^{210}Po} superficial component belonging to the dust that deposited on the acrylic. In fact, the dust removal should cause a reduction of the \ce{^{210}Po} peak component with respect to the tail. Like TPRs before the cleaning, $\text{TPR}_{210Po}^{AC}$ values calculated for each day of measurement keep constant in time. The mean value turns out to be equal to $\text{TPR}_{210Po}^{AC} = (4.6 \pm 0.1)\%$, showing a $(18 \pm 4)\%$ greater fraction with respect to the corresponding value before the cleaning $\text{TPR}_{210Po} = (3.9 \pm 0.1)\%$ (see Table~\ref{tab:7}). 

This TPR increase is smaller than expected, since a compatibility with \ce{^{210}Po} and \ce{^{210}Pb} activity losses percentage previously calculated was expected. However, the peak and tail components defined in Equation~\eqref{eq:1} are discriminated with a simple energy cut in the spectrum, and a contribution due to degraded $\alpha$s emitted is otherwise present also below the peak. These $\alpha$s are emitted by contaminants that are near enough to the edge of the acrylic plate to produce a signal belonging to the peak component, but that are not definitively on the surface. In this way, when the surface is rinsed these \ce{^{210}Po} isotopes are not removed, and the TPR value after the cleaning does not increase as much as the activity decreases.

\subsection{Diffusion Length for \ce{^{210}Po} Implantation}

\begin{figure}
\includegraphics[scale =0.45]{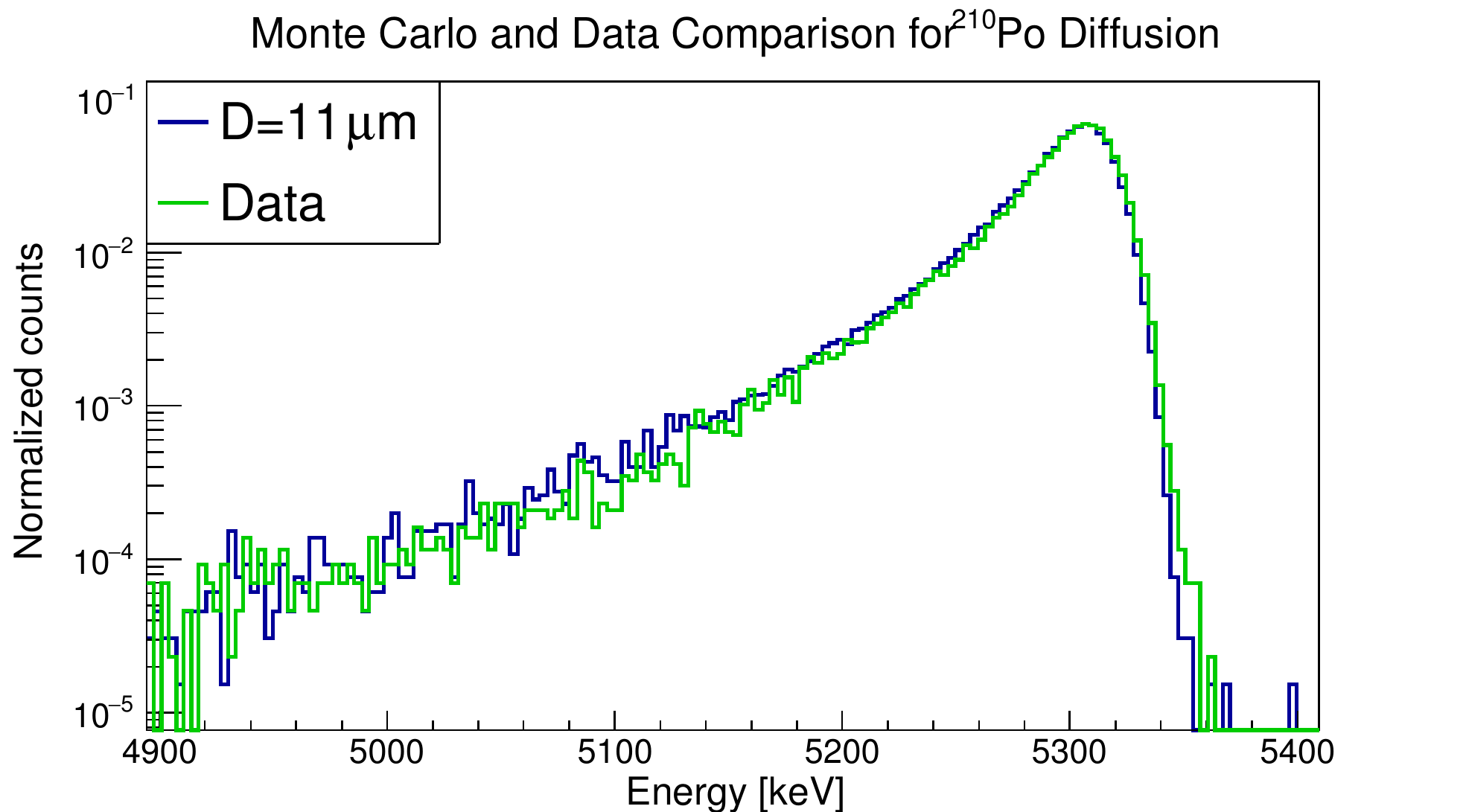}
\caption{Overlap of the simulated \ce{^{210}Po} events (blue line) with $D=\SI{11}{\micro m}$ and the \ce{^{210}Po} data (green line). Both the spectra have been normalized to obtain comparable activity values.}
\label{fig:10}
\end{figure}

In order to evaluate the \ce{^{210}Po} implantation depth, a Monte Carlo simulation has been run with the hypothesis of an exponential diffusion of this isotope inside the acrylic. More precisely, the implantation has been modeled with a function $e^{-x/D}$, where $x$ is the depth and $D$ is the \emph{Diffusion Length}. The TPR values of the simulations at different $D$ values have been compared with the TPR value of the Smooth Acrylic sample after the cleaning, so as to exclude the dust contribution which would have wrongly affected the peak component. The best estimation for the Diffusion Length results to be $D=\SI{11}{\micro m}$, with a TPR for the simulated spectrum equal to $(4.7 \pm 0.1)\%$, compatible with the $\text{TPR}_{210Po}^{AC}$ value of $(4.6 \pm 0.1)\%$. In Figure~\ref{fig:10}, a superimposition of the simulated \ce{^{210}Po} spectrum with $D=\SI{11}{\micro m}$ and the measured \ce{^{210}Po} one is shown.

\subsection{Study of \ce{^{222}Rn} Diffusion} 
\label{RnDiff}

\begin{figure}
\includegraphics[scale =0.22]{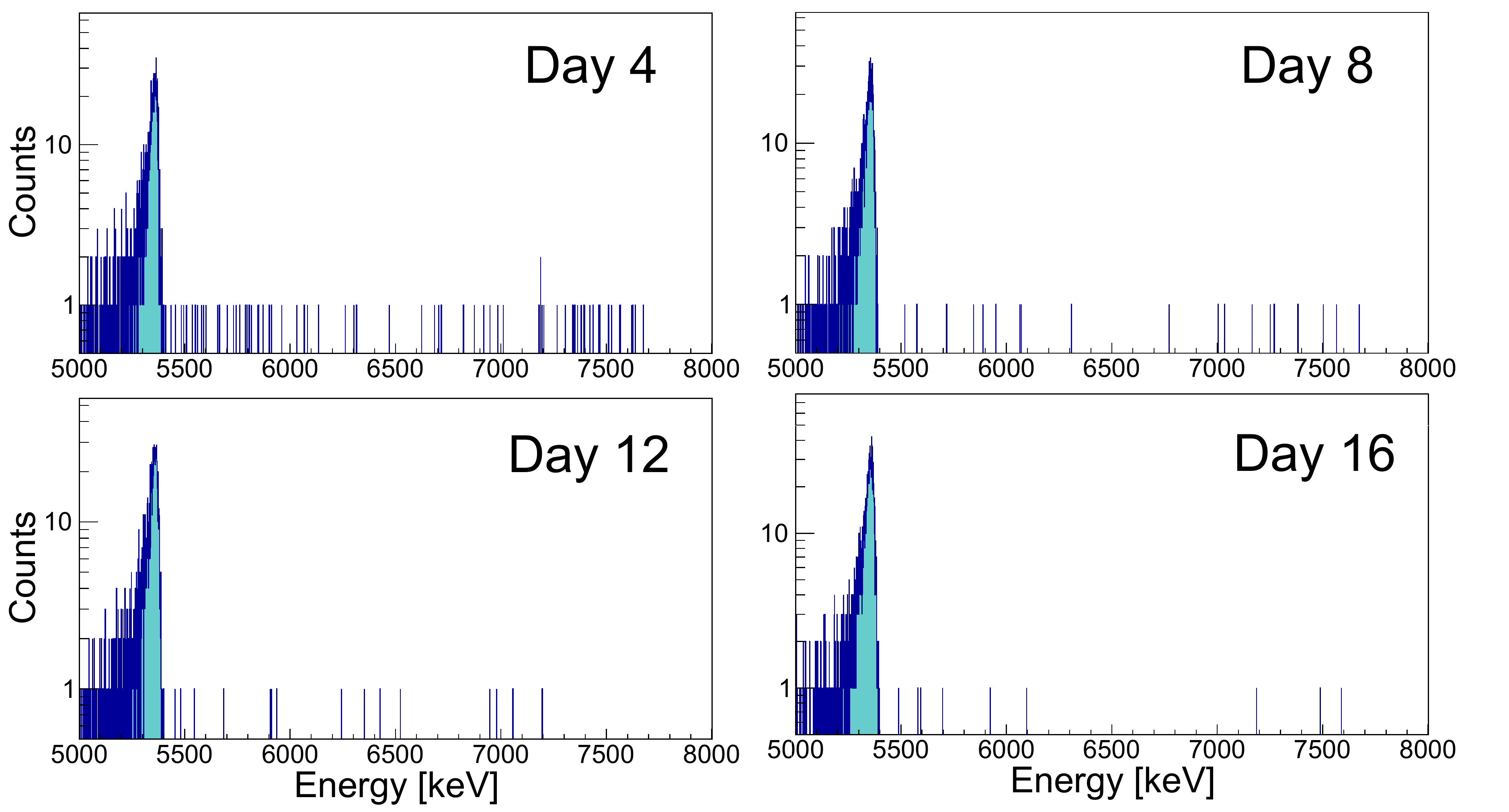}
\caption{Long $\alpha$ measurement spectra in days $4$, $8$, $12$ and $16$ for Opaque Acrylic 2. At \SI{5.3}{MeV} the \ce{^{210}Po} peak is recognizable, and in the energy region on its right a clear decrease of counts is visible.}
\label{fig:SpettriDiff}
\end{figure}

Observing the subsequent daily \ce{^{210}Po} spectra, it has been noticed an evident excess of counts in the energy region from \SIrange{5.4}{7.8}{MeV}, which decreases over time (Fig.~\ref{fig:SpettriDiff}). This energy range includes the peaks of \ce{^{222}Rn} faster daughters \ce{^{218}Po} and \ce{^{214}Po}. A presence of these isotopes several days after the extraction of the samples from the Rn-Box is very unlikely because of their lifetime, unless some radon diffused inside the acrylic during the exposure. In this condition, \ce{^{222}Rn} faster daughters can reach secular equilibrium with their parent inside the plate, and would be still visible after a time frame much greater than their lifetime. More precisely, their decay would follow \ce{^{222}Rn} decay constant ($\lambda = 2.1 \times 10^{-6}$~\si{s^{-1}}), due to the established secular equilibrium condition.

\begin{figure}
\centering
\subfloat
{\includegraphics[scale=0.45]{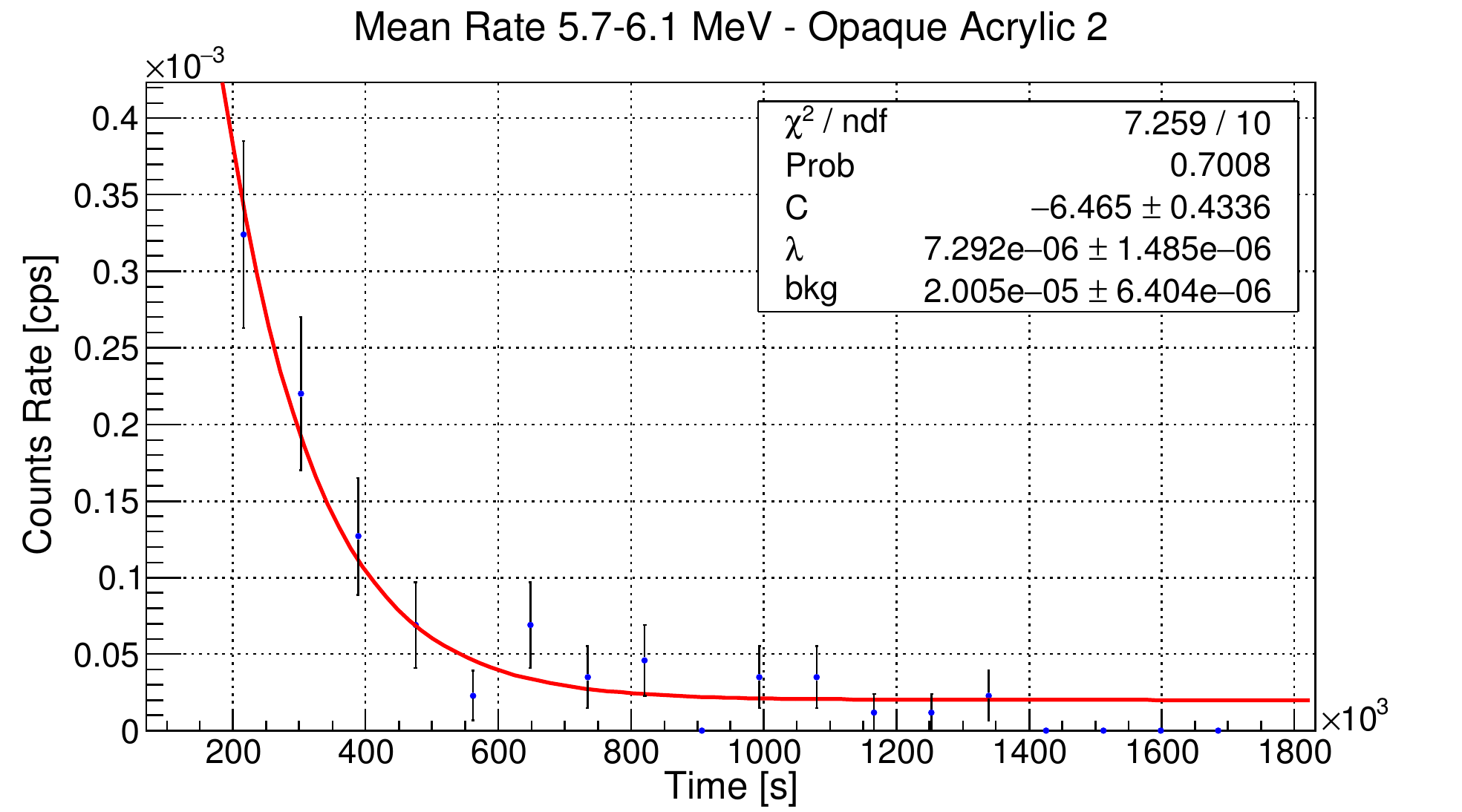}} \\
\subfloat
{\includegraphics[scale=0.45]{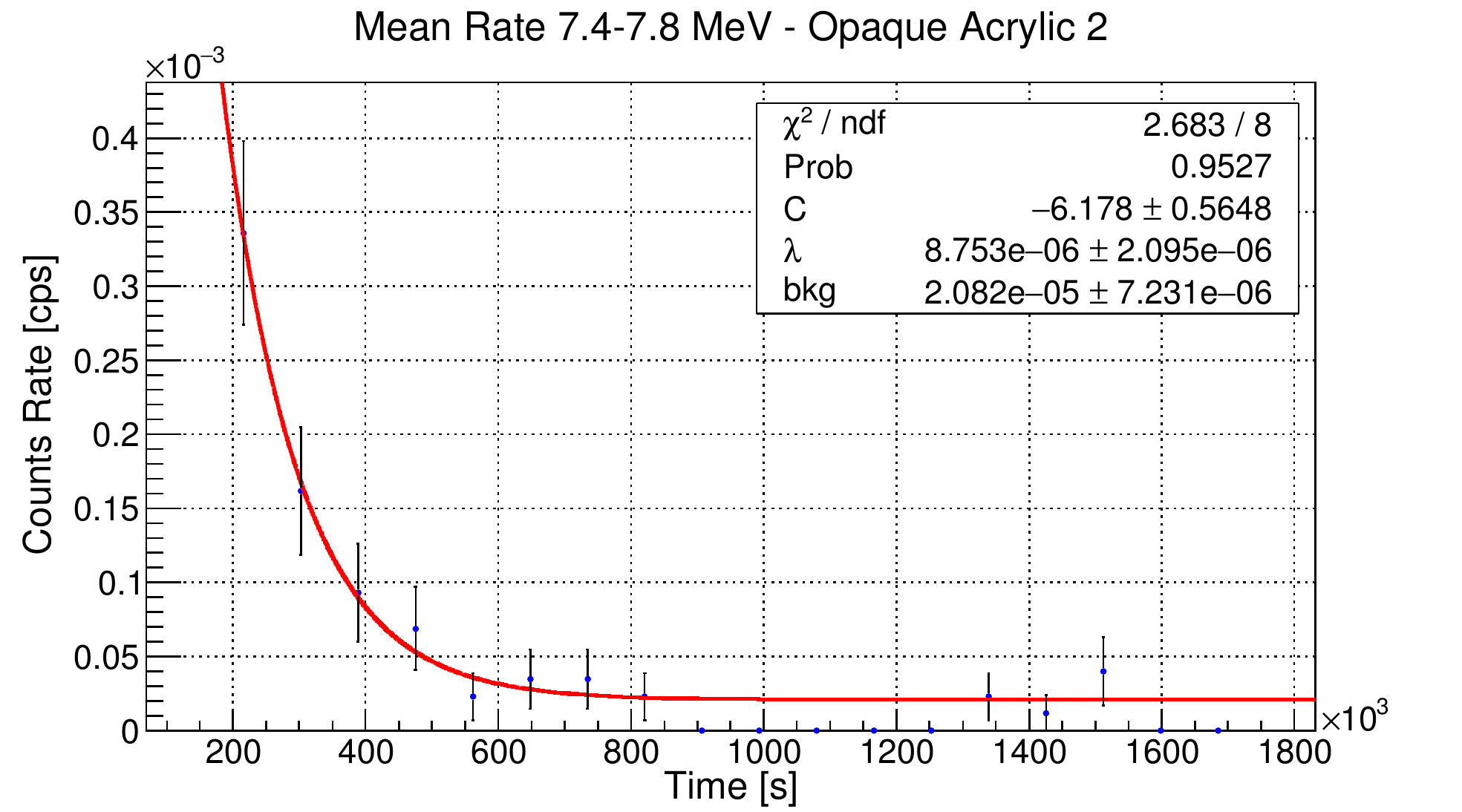}} \quad
\caption{Interpolation of counts rate with an exponential law in the $5.7$-$6.1$~\si{MeV} and $7.4$-$7.8$~\si{MeV} energy ranges. The first two days of measurement are not included.}
\label{fig:DiffRadon}
\end{figure}

In order to verify this hypothesis, each daily spectrum of the \ce{^{210}Po} measurement has been cut in different energy regions, and the counts as a function of time have been fitted with a decreasing exponential law plus a constant. In Figure~\ref{fig:DiffRadon}, this interpolation is shown for the energy ranges of~$5.7$-$6.1$~\si{MeV} and~$7.4$-$7.8$~\si{MeV}, respectively. In order to be sensitive to a signal due to the radon diffusion inside the sample, the first two days of measurement have been excluded from the analysis. In this way, the contribution of the isotopes that have contaminated the sample separately from radon, which would cover the diffusion signal, is not considered. From the analysis, the extrapolated decay constant for the~$5.7$-$6.1$~\si{MeV} range is $\lambda = (7.3 \pm 1.5) \times 10^{-6}$~\si{s^{-1}}, while for the~$7.4$-$7.8$~\si{MeV} range is $\lambda = (8.7 \pm 2.1) \times 10^{-6}$~\si{s^{-1}}. These values are compatible to each other, but do not match with \ce{^{222}Rn} $\lambda = 2.1 \times 10^{-6}$\,\si{s^{-1}}. This result might be explained by the fact that the samples were inside a pumped vacuum chamber during the $\alpha$ measurement; thus, part of \ce{^{222}Rn} (reasonably the most superficial component) diffused out of them, following the inverted concentration gradient due to the void in the vacuum chamber. This assertion would also explain why there is not an evident excess of counts at \SI{5.5}{MeV} (see Sec.~\ref{sec:3}), where \ce{^{222}Rn} alpha peak should be present. This result is anyway not sufficient to come to a conclusion about \ce{^{222}Rn} diffusion inside acrylic.

\begin{figure}
\includegraphics[scale =0.45]{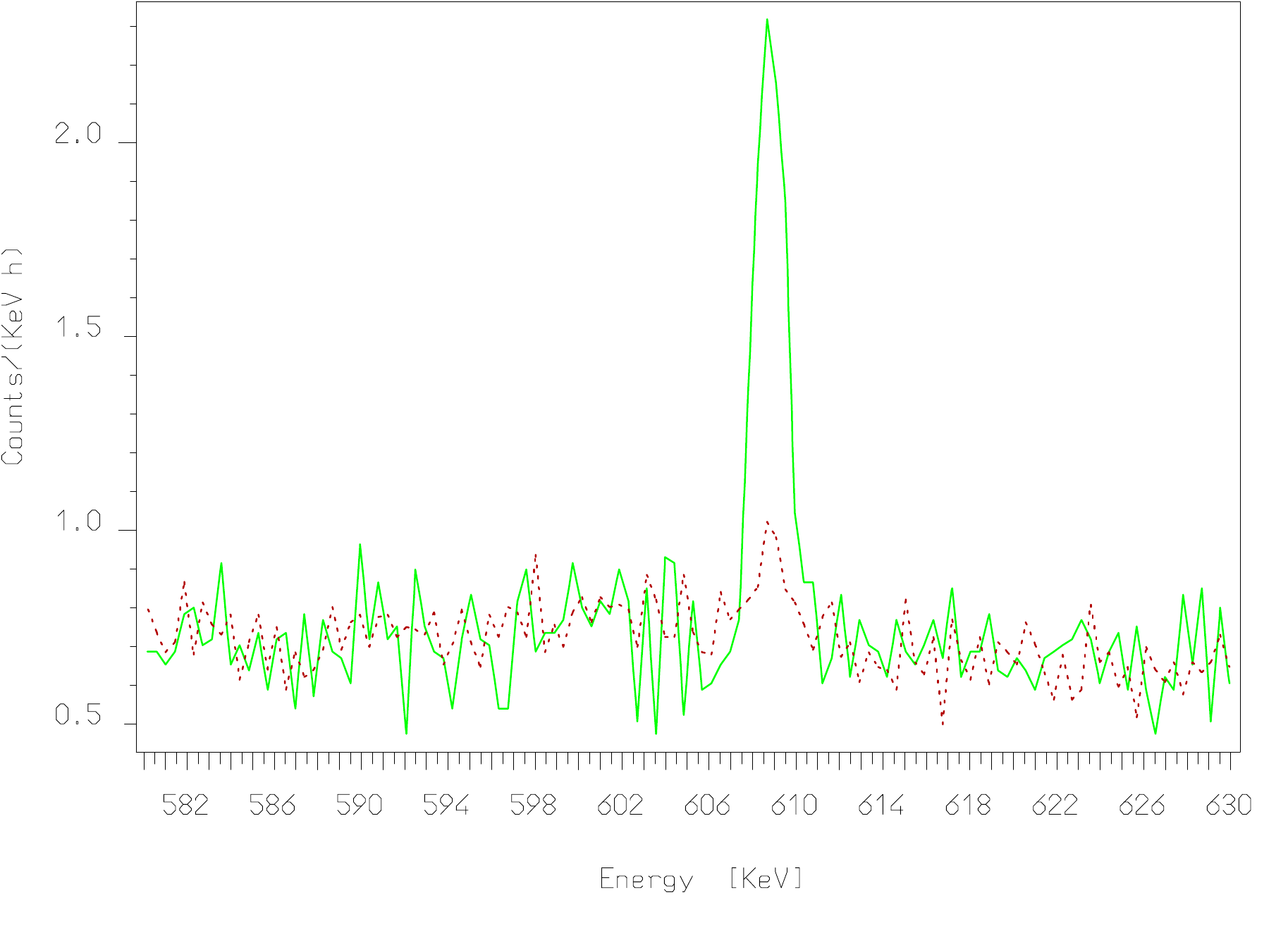}
\caption{Overlap of \ce{^{214}Bi} signal (green solid line) and the  background of the $\gamma$-ray detector (brown dashed line). An excess of counts at \SI{609}{keV} is clearly distinguishable.}
\label{GammaDiffBackg}
\end{figure}

\begin{figure}
\includegraphics[scale =0.45]{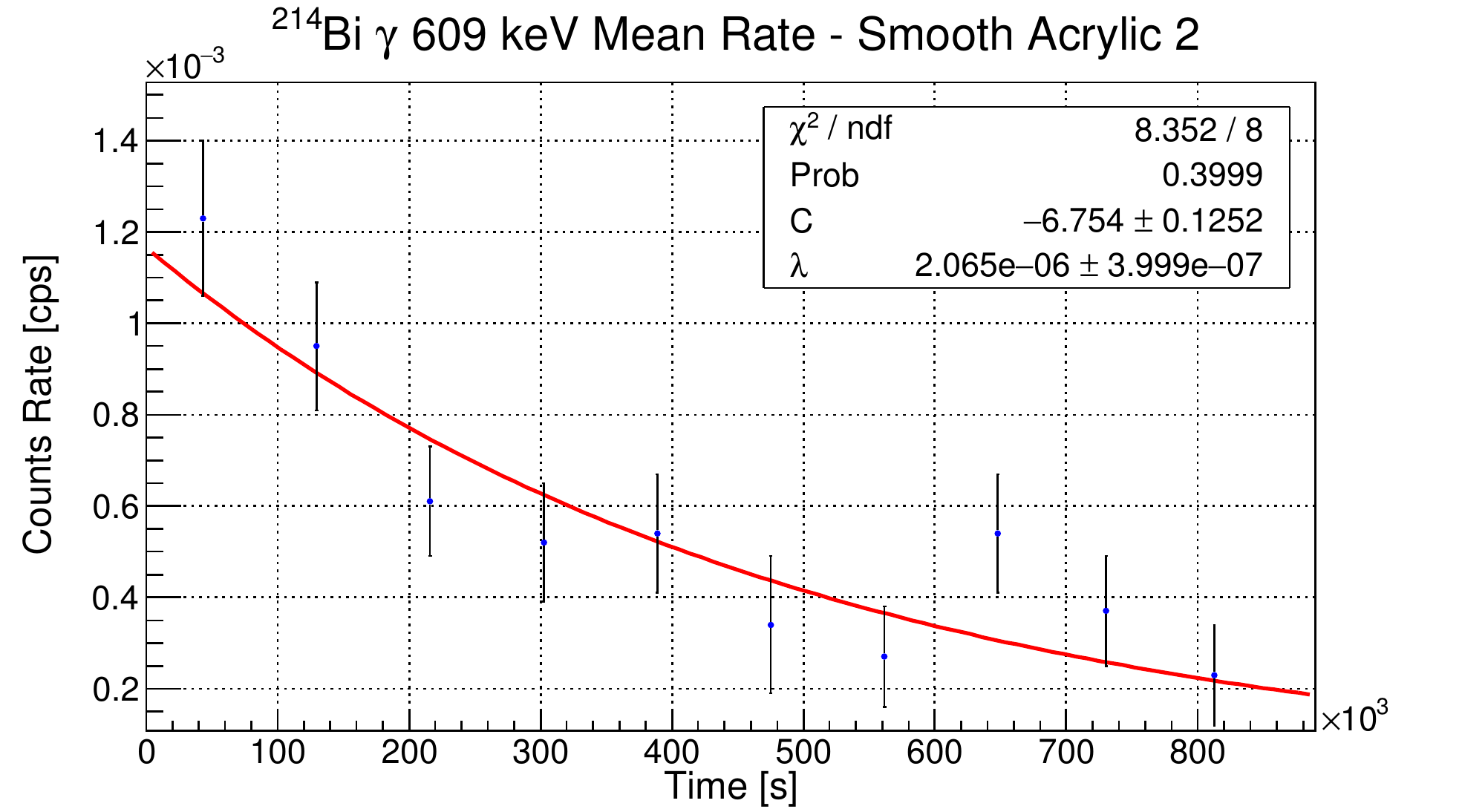}
\caption{Interpolation of \ce{^{214}Bi} activity from \SI{609}{keV} gamma line over time. The decay constant $\lambda$ is in accord with \ce{^{222}Rn} one.}
\label{fig:GammaDiff}
\end{figure}

As a further study, a new smooth sample (Smooth Acrylic 2), with an exposure time of $277$ days, has been extracted from the Rn-Box and new $\gamma$ measurements of \ce{^{214}Bi} \SI{609}{keV} line have been carried out using the HPGe detector, where no vacuum is needed. These measurements have been performed one day after the extraction of the sample, so as to establish secular equilibrium between \ce{^{214}Bi} and \ce{^{222}Rn}. Also in this case, an excess of \ce{^{214}Bi} counts and their exponential decrease over time are evident, as it is shown in Figures~\ref{GammaDiffBackg} and~\ref{fig:GammaDiff}. For these measurements, the decay constant $\lambda = (2.1 \pm 0.4) \times 10^{-6}$~\si{s^{-1}} from the interpolation is compatible with \ce{^{222}Rn} one, as expected.

Thanks to this result, the diffusion of \ce{^{222}Rn} inside the acrylic samples due to a concentration gradient is not to be totally excluded. However, it is necessary to perform further investigation to verify this phenomenon.

\section{Conclusion}

This study made possible to understand the mechanisms of acrylic exposure to radon-222. 

Thanks to a complete model of the time evolution of \ce{^{218}Po} and \ce{^{214}Po} activity, it has been possible to state that alpha decay is the only responsible of a deep implantation of a nuclide inside acrylic. This assertion has been ulteriorly confirmed by the study of \ce{^{210}Po} Tail-to-Peak Ratio. $\text{TPR}_{210Po}$ resulted to be significantly greater than the TPR values estimated for its alpha emitters progenitors \ce{^{214}Po} and \ce{^{218}Po}, proving that the deeper implantation of this isotope is caused by subsequent alpha decays. A clear difference of the TPR values between smooth and opaque samples highlighted the dependency of the contamination depth by the acrylic texture. In particular, the opaque texture is more subjected to a deeper contamination than the smooth one. To quantify the implantation depth of the radiocontaminants, the data have been compared with a Monte Carlo simulation of a \ce{^{210}Po} source exponentially diffused inside the acrylic. The best accord with the data has been obtained for a Diffusion Length of \SI{11}{\micro m}. With $\gamma$ measurements, both \ce{^{210}Pb} superficial and deep contamination components have been quantified. The linear growth of \ce{^{210}Pb} activity over time has been also proven. Moreover, \ce{^{222}Rn} diffusion inside the samples has been observed thanks to dedicated \ce{^{214}Bi} gamma measurements, although further investigation is necessary to fully understand the features of this contamination mechanism. 

The positive results of this work proved the feasibility of the proposed approach in a general study of radioactive contamination of materials. The methods here validated can be applied in further studies of materials contaminated by different isotopes.

\end{document}